\documentclass[aip,jcp,amsmath,amssymb,
runinaddress,
reprint]
{revtex4-2}

\usepackage{graphicx}
\usepackage{dcolumn}
\usepackage{bm}
\usepackage{mathptmx}
\usepackage{booktabs}

\usepackage{color}
\usepackage[colorlinks=true,
linkcolor=blue,
urlcolor=blue,
citecolor=blue,
anchorcolor=blue]
{hyperref}

\newcommand{\dd}{\mathrm{d}}
\newcommand{\e}{\mathrm{e}}
\newcommand{\mbr}{\mathbf{r}}
\newcommand{\mbR}{\mathbf{R}}

\newcolumntype{d}[1]{D{.}{.}{#1}}

\begin{document}

\title{Third and fourth density and acoustic virial coefficients of neon from first-principles calculations}

\author{Robert \surname{Hellmann}}
\email{robert.hellmann@hsu-hh.de}
\affiliation{Institut f\"ur Thermodynamik, Helmut-Schmidt-Universit\"at / Universit\"at der Bundeswehr Hamburg, Holstenhofweg~85, 22043 Hamburg, Germany}
\author{Giovanni \surname{Garberoglio}}
\email{garberoglio@ectstar.eu}
\affiliation{European Centre for Theoretical Studies in Nuclear Physics and Related Areas (FBK-ECT*), Strada delle Tabarelle 286, 38123 Trento, Italy}

\date{\today}


\begin{abstract}

The third and fourth density and acoustic virial coefficients of neon were determined at temperatures between 10 and 5000\;K from first principles employing the path-integral Monte Carlo (PIMC) approach. For these calculations, we used the pair potential of Hellmann \emph{et~al.}\ [J.\ Chem.\ Phys.\ \textbf{154}, 164304 (2021)], which is based on supermolecular \emph{ab initio} calculations with basis sets of up to octuple-zeta quality and levels of theory up to coupled cluster with single, double, triple, quadruple, and perturbative pentuple excitations \mbox{[CCSDTQ(P)]}. The potential also accounts for relativistic, retardation, and post-Born--Oppenheimer effects and is provided with reliable uncertainty estimates. To incorporate nonadditive interactions, we developed a nonadditive three-body potential based on extensive supermolecular \mbox{CCSD(T)}, \mbox{CCSDT}, and \mbox{CCSDT(Q)} calculations with basis sets of up to sextuple-zeta quality. This potential also accounts for relativistic effects. The very small nonadditive four-body contributions to the fourth virial coefficients were considered using a relatively simple nonadditive four-body potential based on supermolecular \mbox{CCSD(T)} calculations. We calculated the third and fourth density and third acoustic virial coefficients directly by PIMC and the fourth acoustic virial coefficient indirectly using thermodynamic relations between the density and acoustic virial coefficients. The uncertainties of the pair potential and those estimated for our nonadditive three-body potential were rigorously propagated in the PIMC calculations into uncertainties for the virial coefficients. These uncertainties are distinctly smaller than those of almost all of the corresponding experimental virial coefficient data.


\end{abstract}




\maketitle


\section{Introduction}
\label{sec:1}


Noble gases are the simplest and most important model substances for studying thermophysical properties of fluids both theoretically and experimentally. They are chemically inert, making them easy to handle in experiments, and their monatomic nature and spherical symmetry allow the use of essentially exact statistical-mechanical approaches in combination with accurate models of the interatomic interactions for the calculation of their properties. These features also make them ideal working gases in gas-based temperature and pressure metrology, where the thermophysical properties have to be known with extremely low uncertainties.\cite{Garberoglio:2023}

For helium, interatomic interactions can be studied so accurately with quantum-chemical \emph{ab initio} approaches today that the resulting thermophysical properties have uncertainties that are far lower than those achievable experimentally.\cite{Czachorowski:2020,Garberoglio:2020,Gokul:2021,Garberoglio:2021b,Lang:2023,Wheatley:2023,Garberoglio:2023,Binosi:2024,Garberoglio:2024,Guo:2026} This is one of the key reasons why helium (more precisely the $^4$He isotope) is the most widely used working gas in gas metrology.\cite{Moldover:2014,Pitre:2017,Gaiser:2017,Fischer:2018,Gavioso:2019,Gaiser:2020,Gaiser:2020a,Ripa:2021,Gaiser:2021b,Gaiser:2022,Garberoglio:2023} In addition, helium is quite cheap and commercially available with very high purity. Argon is also commonly used in gas metrology,\cite{Moldover:2014,Underwood:2017,Fischer:2018,Gaiser:2020a,Garberoglio:2023} and there is growing interest in using neon for such applications as well.\cite{Gaiser:2017,Gaiser:2020a,Ripa:2021,Garberoglio:2023} The advantage of both neon and argon over helium is that their higher masses and polarizabilities make the experiments less sensitive to impurities. A disadvantage of both gases is that they are usually not available as pure isotopes, so that samples from different sources can have measurable differences in certain thermophysical properties, especially the speed of sound, due to variations in isotopic composition. Moreover, the thermophysical properties of neon and argon cannot be predicted from first principles as accurately as those of helium.\cite{Garberoglio:2023} This is due to the fact that neon and argon atoms have far more electrons than helium atoms, which limits the accuracy with which interatomic properties, such as the pair potential or the induced pair polarizability, can be obtained with quantum-chemical \emph{ab initio} approaches. A disadvantage for neon is also its high price. However, the thermophysical properties of neon can be calculated more accurately than those of argon due to its fewer electrons, and in a recent paper Hellmann \emph{et~al.}\cite{Hellmann:2021} presented a state-of-the-art \emph{ab initio} pair potential and induced pair polarizability with which they calculated thermophysical properties of low-density neon gas with uncertainties that are small enough for applications in gas metrology.

An accurate description of the thermodynamic properties of neon gas beyond the low-density regime covered by the work of Hellmann \emph{et~al.}\cite{Hellmann:2021} is also of significant relevance for gas metrology. This requires considering not only pair interactions (as in the paper of Hellmann \emph{et~al.}) but also multi-body interactions. The latter cannot be correctly described solely by pair potentials as, for example, the interaction energy of a triplet of neon atoms differs slightly from the sum of the corresponding three pair interaction energies, with the difference being the so-called nonadditive three-body interaction energy. The concept of nonadditivity can be systematically expressed in the form of an exact many-body expansion of the total interaction potential $V_N$ of an $N$-atom system,
\begin{equation}
\label{manybody}
V_N = \sum_{i<j}V_{ij} + \sum_{i<j<k}\Delta V_{ijk} + \sum_{i<j<k<l}\Delta V_{ijkl} + \ldots,
\end{equation}
where $V_{ij}$ is the pair potential between atoms $i$ and $j$, $\Delta V_{ijk}$ denotes the nonadditive three-body potential between atoms $i$, $j$, and $k$,
\begin{equation}
\label{dVijk_def}
\Delta V_{ijk} = V_{ijk} - V_{ij} - V_{ik} - V_{jk},
\end{equation}
with $V_{ijk}$ being the total interaction potential between the three atoms, and $\Delta V_{ijkl}$ is the nonadditive four-body potential between atoms $i$, $j$, $k$, and $l$,
\begin{align}
\label{dVijkl_def}
\Delta V_{ijkl} ={}& V_{ijkl} - V_{ij} - V_{ik} - V_{il} - V_{jk} - V_{jl} - V_{kl} \nonumber\\
& - \Delta V_{ijk} - \Delta V_{ijl} - \Delta V_{ikl} - \Delta V_{jkl},
\end{align}
with $V_{ijkl}$ being the total interaction potential between the four atoms. For helium, the many-body expansion has been found to converge rapidly.\cite{Wheatley:2023}

In gas-based temperature and pressure metrology, the equation of state of the working gas is almost always expressed in the form of a virial expansion. The expansion of pressure $p$ in powers of molar density $\rho_\text{m}$ is given by
\begin{equation}
\label{eq:Virial_expansion_density}
\frac{p}{\rho_\mathrm{m} RT} = 1 + B(T)\rho_\mathrm{m} + C(T)\rho_\mathrm{m}^2 + D(T)\rho_\mathrm{m}^3 + \ldots,
\end{equation}
where $R$ is the molar gas constant, $T$ is the temperature, and $B$, $C$, and $D$ are the second, third, and fourth density virial coefficients, which depend only on temperature. Essentially exact statistical-mechanical expressions are available that relate the density virial coefficients to the interaction potentials. The $n$th density virial coefficient depends on the interactions within isolated groups of up to $n$ particles, so that knowledge of the pair potential is sufficient to calculate the second density virial coefficient but not the higher ones. Virial expansions can be formulated for various thermodynamic properties. The expansion for the speed of sound $w$ in powers of pressure, which is used in acoustic gas thermometry, is usually written in the form\cite{Gillis:1996,Garberoglio:2023}
\begin{equation}
\label{eq:Virial_expansion_acoustic}
w^2 = \frac{\gamma^{\circ}RT}{M}\left[1 + \frac{\beta_\text{a}(T)}{RT}p + \frac{\gamma_\text{a}(T)}{RT}p^2 + \frac{\delta_\text{a}(T)}{RT}p^3 + \ldots\right],
\end{equation}
where $\gamma^{\circ}=c_{p}^{\circ}/c_{V}^{\circ}$ is the ratio of the isobaric and isochoric ideal gas heat capacities (with $\gamma^{\circ}=5/3$ for noble gases), $M$ is the average molar mass of the gas (for natural neon \mbox{$M=20.1797$\;g/mol\cite{Laeter:2003}}), and $\beta_\text{a}$, $\gamma_\text{a}$, and $\delta_\text{a}$ are the second, third, and fourth acoustic virial coefficients. While $\beta_\text{a}$ and $B$ have the same dimensions, $\gamma_\text{a}$ and $\delta_\text{a}$ are usually quoted as \mbox{$RT\gamma_\text{a}$} and \mbox{$(RT)^2\delta_\text{a}$} to obtain values with the same dimensions as $C$ and $D$, respectively. The $n$th acoustic virial coefficient can be obtained from the second to $n$th density virial coefficients and their first and second temperature derivatives.\cite{Gillis:1996,Garberoglio:2023}

Hellmann \emph{et~al.}\cite{Hellmann:2021} already provided the second density and acoustic virial coefficients for both $^{20}$Ne and the natural isotopic composition as found in the atmosphere, which is 90.48\% $^{20}$Ne, 0.27\% $^{21}$Ne, and 9.25\% $^{22}$Ne.\cite{Laeter:2003} Note that the $\beta_\text{a}$ values for $^{20}$Ne and natural neon given by Hellmann \emph{et~al.}\ were accidentally swapped at temperatures from 7 to 28\;K. The differences between the two sets of $\beta_\text{a}$ values are very small and within the mutual uncertainties. In the \textcolor{blue}{supplementary material}, we provide a corrected version of the respective table from the supplementary material of the paper of Hellmann \emph{et~al.}

\begin{figure}
\includegraphics{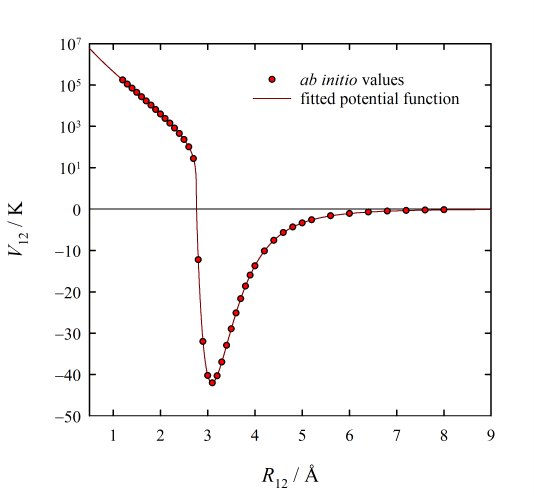}
\caption{\emph{Ab initio} neon pair potential as reported by Hellmann \emph{et~al.}\cite{Hellmann:2021}}
\label{fig:V12}
\end{figure}

In this work, we extend the study of Hellmann \emph{et~al.}\cite{Hellmann:2021} to the third and fourth density and third and fourth acoustic virial coefficients. We reuse the \emph{ab initio} pair potential developed by Hellmann \emph{et~al.}\ for this purpose. It is given for atom-atom separations $R_{12}\geqslant1.2$\;\AA{} by a modified Tang--Toennies function,\cite{Tang:1984}
\begin{align}
\label{eq:Vfit}
  V_{12}\left(R_{12}\right) ={}& A\exp\left(a_{1} R_{12} + a_{2} R_{12}^{2} + a_{-1} R_{12}^{-1} + a_{-2} R_{12}^{-2}\right) \nonumber\\ & - \sum_{n=3}^{8}f_{2n}\left(R_{12}\right)\frac{C_{2n}}{R_{12}^{2n}}\left[1 - \exp\left(-bR_{12}\right)\sum_{k=0}^{2n}\frac{\left(bR_{12}\right)^k}{k!}\right],
\end{align}
where $f_{2n}(R_{12})$ are retardation functions given for $n=3$ and $n=4$ by
\begin{equation}
  f_{2n}(R_{12}) = \frac{g_{2n}(R_{12})}{g_{2n}(R_{12})+R_{12}},
\end{equation}
with
\begin{equation}
  g_{2n}(R_{12}) = d_{2n,1}+d_{2n,2}\exp\left\{\sum_{k=1}^{5}d_{2n,k+2}\left[\ln\left(R_{12}/\AA{}\right)\right]^k\right\},
\end{equation}
and for $n>4$ by \mbox{$f_{2n}(R_{12})=1$}. Hellmann \emph{et~al.}\ fitted the potential function to interaction energies for 42 separations from $1.2$ to $8.0$\;\AA{} obtained from counterpoise-corrected\cite{Boys:1970} supermolecular \emph{ab initio} calculations with basis sets of up to octuple-zeta quality supplemented by bond functions and at levels of theory up to coupled cluster with single, double, triple, quadruple, and perturbative pentuple excitations \mbox{[CCSDTQ(P)]}.\cite{Kallay:2005} Corrections for relativistic and post-Born--Oppenheimer effects were also computed. The retardation corrections in the form of the $f_{6}(R_{12})$ and $f_{8}(R_{12})$ functions were determined separately from these quantum-chemical calculations and applied after a fit of the potential function without retardation. For \mbox{$R_{12}<1.2$\;\AA{}}, a simplified potential $V_\mathrm{12,sr}(R_{12})$ given by
\begin{equation}
\label{eq:Vfitsmall}
  V_\mathrm{12,sr}(R_{12}) = \frac{A_\mathrm{sr}}{R_{12}}\exp\left(-a_\mathrm{sr}R_{12}\right)
\end{equation}
was fitted by Hellmann \emph{et~al.}\ such that $V_\mathrm{12,sr}$ and its first derivative at $R_{12}=1.2$\;\AA{} match the respective values for $V_{12}$. The parameters $A$, $a_{1}$, $a_{2}$, $a_{-1}$, $a_{-2}$, $b$, $C_{6}$, $C_{8}$, $C_{10}$, $C_{12}$, $C_{14}$, $C_{16}$, $d_{6,1}$ to $d_{6,7}$, $d_{8,1}$ to $d_{8,7}$, $A_\mathrm{sr}$, and $a_\mathrm{sr}$ are given in the original publication and are not repeated here. The pair potential of Hellmann \emph{et~al.}\ is depicted in Fig.~\ref{fig:V12}, where $V_{12}$ is given in kelvin, i.e., we divided $V_{12}$ by the Boltzmann constant $k_\mathrm{B}$ but omit $k_\mathrm{B}$ from the notation for brevity. Hellmann \emph{et~al.}\ also provided pair potentials fitted to the \emph{ab initio} values shifted either all upwards or all downwards by their estimated standard uncertainties. These two additional potentials are required for obtaining the uncertainty contributions to the virial coefficients due to the pair potential.

To describe the nonadditive contributions to the virial coefficients, we developed new \emph{ab initio} nonadditive three-body and four-body potentials, which are presented in Secs.~\ref{sec:2} and \ref{sec:3}, respectively. We calculated the virial coefficients at temperatures from 10 to 5000\;K fully quantum-mechanically, noting that we obtained the fourth acoustic virial coefficient in an indirect manner using thermodynamic relations between the density and acoustic virial coefficients.\cite{Gillis:1996,Garberoglio:2023} The details of these calculations are given in Sec.~\ref{sec:4}. The calculated virial coefficients are compared with experimental data and previous first-principles results in Sec.~\ref{sec:5}, followed by conclusions in Sec.~\ref{sec:6}.


\section{Nonadditive three-body potential}
\label{sec:2}



\subsection{\emph{Ab initio} calculations}
\label{sec:2.1}

\begin{figure}
\includegraphics{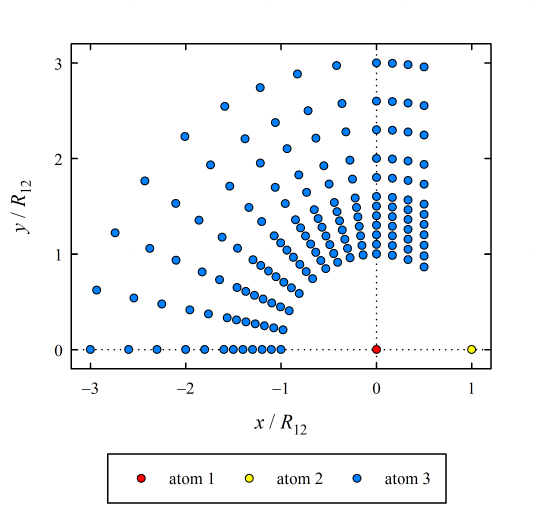}
\caption{The 156 triangular shapes considered for the quantum-chemical \emph{ab initio} calculations of nonadditive three-body interaction energies, shown here by the respective positions of neon atom 3 relative to the positions of neon atoms 1 and 2, whose positions were arbitrarily chosen to be at the origin and on the positive $x$ axis, respectively.}
\label{fig:triangles}
\end{figure}

To adequately sample the three-dimensional nonadditive three-body potential energy surface by quantum-chemical \emph{ab initio} calculations, a large set of grid points has to be set up. As the coordinates for this grid, we chose the interatomic separations $R_{12}$ and $R_{13}$ and the interior angle at atom 1 of the triangle formed by the three atoms, $\theta_1$, with the atoms being labeled such that $R_{12}\leqslant R_{13}\leqslant R_{23}$. The grid points were then obtained as follows: First, we selected 156 unique triangular shapes defined by the ratio \mbox{$R_{13}/R_{12}$} and the angle $\theta_1$. These shapes are visualized in Fig.~\ref{fig:triangles}.
In the next step, 23 $R_{12}$ values from 1.4 to 10.0\;\AA{} were considered for each shape, resulting in \mbox{$156\times23=3588$} configurations, of which those with \mbox{$R_{12}+R_{13}+R_{23}>30$\;\AA{}} were discarded. This resulted in a final grid of 2748 points for the \emph{ab initio} calculations.

The nonadditive three-body interaction energies $\Delta V_{123}$ were determined with the supermolecular approach, which relates interaction energies to differences in total electronic energies,
\begin{align}
\label{eq:supermolecule3}
\Delta V_{123} ={}& V_{123} - V_{12} - V_{13} - V_{23} \nonumber\\
={}& E_{123} - E_{1} - E_{2} - E_{3} - \left(E_{12}-E_{1}-E_{2}\right) \nonumber\\
& - \left(E_{13}-E_{1}-E_{3}\right) - \left(E_{23}-E_{2}-E_{3}\right) \nonumber\\
={}& E_{123} - E_{12} - E_{13} - E_{23} + E_{1} + E_{2} + E_{3},
\end{align}
where $E_{123}$ is the total electronic energy of the three interacting atoms, $E_{ij}$ is the corresponding quantity for an isolated pair of atoms $i$ and $j$ with the same separation $R_{ij}$ as in the three-atom system, and $E_{i}$ is the total electronic energy of an isolated atom. To avoid the basis set superposition error (BSSE), the counterpoise procedure\cite{Boys:1970} was applied, i.e., all electronic energies were obtained in the full basis set of the three-atom system. We used the \mbox{CFOUR} program\cite{cfour,Matthews:2020} to perform these calculations.

For all grid points, $\Delta V_{123}$ was calculated as a sum of six terms,
\begin{align}
\label{eq:dV123_abinitio}
\Delta V_{123} ={}& \Delta V_{123}^\mathrm{SCF} + \Delta V_{123}^\mathrm{CCSD(T)c} + \Delta V_{123}^\mathrm{T-(T)} + \Delta V_{123}^\mathrm{(Q)-T} \nonumber\\
& + \Delta V_{123}^\mathrm{core} + \Delta V_{123}^\mathrm{rel},
\end{align}
where $\Delta V_{123}^\mathrm{SCF}$ is the Hartree--Fock self-consistent-field contribution, $\Delta V_{123}^\mathrm{CCSD(T)c}$ is the correlation energy contribution at the \mbox{CCSD(T)}\cite{Raghavachari:1989} level of theory within the frozen-core (FC) approximation, $\Delta V_{123}^\mathrm{T-(T)}$ and $\Delta V_{123}^\mathrm{(Q)-T}$ are corrections for the \mbox{FC-CCSDT\cite{Noga:1987}} and \mbox{FC-CCSDT(Q)\cite{Bomble:2005}} levels of theory, respectively, $\Delta V_{123}^\mathrm{core}$ is a correction for core-core and core-valence correlation effects, and $\Delta V_{123}^\mathrm{rel}$ is a correction for scalar relativistic effects.

The contributions $\Delta V_{123}^\mathrm{SCF}$ and $\Delta V_{123}^\mathrm{CCSD(T)c}$ were obtained together using the \mbox{d-aug-cc-pV$X$Z}\cite{Dunning:1989,Kendall:1992,Wilson:1996,vanMourik:1999} basis sets up to sextuple-zeta quality (\mbox{$X=6$}). The results for $\Delta V_{123}^\mathrm{CCSD(T)c}$ at the two highest basis set levels were used to extrapolate this contribution to the complete basis set (CBS) limit with the $X^{-3}$ scheme,\cite{Helgaker:1997,Halkier:1998}
\begin{equation}
\label{CBS_1}
\Delta V_{123}^\mathrm{CCSD(T)c/CBS} = \frac{6^3 \Delta V_{123}^{\mathrm{CCSD(T)c}/X=6}-5^3 \Delta V_{123}^{\mathrm{CCSD(T)c}/X=5}}{6^3-5^3},
\end{equation}
while the SCF contributions were not extrapolated because they converge more rapidly to the CBS limit and are essentially converged at the \mbox{$X=6$} basis set level. However, at this highest basis set level it was often not possible to obtain the desired very tight convergence thresholds for the SCF iterations, which affected mostly the SCF contribution itself but in some cases also significantly the \mbox{CCSD(T)} correlation contribution. The resulting numerical noise is most evident and problematic for large triangles of all shapes, where the absolute values of $\Delta V_{123}$ are very small.

We divided the 2748 grid points for which $\Delta V_{123}^\mathrm{SCF}$ and $\Delta V_{123}^\mathrm{CCSD(T)c}$ were calculated into four groups, with a different way of dealing with the numerical issues for each group. The first group are 1354 points for which no further action was taken because the numerical issues were assessed to be unproblematic. The second group are 87 points for which $\Delta V_{123}^\mathrm{SCF}$ obtained at the \mbox{$X=6$} level was assessed (based on the observed convergence pattern from \mbox{$X=2$} to \mbox{$X=5$}) to be less accurate due to numerical noise than the value obtained at the \mbox{$X=5$} level. Therefore, for these points, $\Delta V_{123}^\mathrm{SCF}$ at the \mbox{$X=5$} basis set level was used together with the $\Delta V_{123}^\mathrm{CCSD(T)c}$ values obtained from the CBS extrapolation with \mbox{$X=5$} and \mbox{$X=6$}. The third group are 1109 points for which the true absolute value of the SCF contribution should be so small that setting $\Delta V_{123}^\mathrm{SCF}$ to zero was assessed to be more accurate than using the calculated value at any basis set level because the calculated values always start to unphysically oscillate around zero due to numerical noise once the size of the triangle of atoms becomes large enough. The \mbox{CCSD(T)} correlation contributions for these points were calculated in the same manner as for those of the first and second group. The fourth group are 198 points for which $\Delta V_{123}^\mathrm{CCSD(T)c}$ at the \mbox{$X=6$} level differed in an unphysical manner from the observed convergence pattern of the respective values for \mbox{$X<6$} due to numerical noise. These 198 points were discarded.

The corrections for the higher \mbox{CCSDT} level of theory, $\Delta V_{123}^\mathrm{T-(T)}$, were calculated as the differences between \mbox{CCSDT} and \mbox{CCSD(T)} interaction energies obtained within the FC approximation using the \mbox{d-aug-cc-pV$X$Z} basis sets with \mbox{$X=2$~(D)} and \mbox{$X=3$~(T)}. Similarly, the corrections for the \mbox{CCSDT(Q)} level of theory, $\Delta V_{123}^\mathrm{(Q)-T}$, were determined as the differences between \mbox{CCSDT(Q)} and \mbox{CCSDT} interaction energies calculated within the FC approximation but using the smaller \mbox{aug-cc-pV$X$Z} basis sets with \mbox{$X=2$} and \mbox{$X=3$}. The resulting values for the corrections were extrapolated to their CBS limits with the $X^{-3}$ scheme,
\begin{align}
\label{CBS_2}
\Delta V_{123}^\mathrm{T-(T)/CBS} ={}& \frac{3^3 \Delta V_{123}^{\mathrm{T-(T)}/X=3}-2^3 \Delta V_{123}^{\mathrm{T-(T)}/X=2}}{3^3-2^3},\\
\Delta V_{123}^\mathrm{(Q)-T/CBS} ={}& \frac{3^3 \Delta V_{123}^{\mathrm{(Q)-T}/X=3}-2^3 \Delta V_{123}^{\mathrm{(Q)-T}/X=2}}{3^3-2^3}.
\end{align}
These CBS estimates are of course not as reliable as those for $\Delta V_{123}^\mathrm{CCSD(T)c}$ due to the much smaller basis sets that had to be used with the very costly \mbox{CCSDT} and \mbox{CCSDT(Q)} methods. Of the two corrections, $\Delta V_{123}^\mathrm{T-(T)}$ is the more important one, being typically a few times larger than $\Delta V_{123}^\mathrm{(Q)-T}$.

The corrections for core-core and core-valence correlation effects, $\Delta V_{123}^\mathrm{core}$, were calculated by taking the differences between \mbox{CCSD(T)} interaction energies obtained with and without the FC approximation using the \mbox{d-aug-cc-pwCV$X$Z}\cite{Dunning:1989,Kendall:1992,Peterson:2002} basis sets with $X=2$ and $X=3$. The contributions of scalar relativistic effects, $\Delta V_{123}^\mathrm{rel}$, were approximated by the second-order direct perturbation theory (DPT2)\cite{Kutzelnigg:1995,Klopper:1997} corrections to interaction energies calculated at the FC-CCSD(T) level of theory using uncontracted \mbox{d-aug-cc-pV$X$Z} basis sets with $X=3$ and $X=4$. The values obtained for $\Delta V_{123}^\mathrm{core}$ and $\Delta V_{123}^\mathrm{rel}$ were also extrapolated to their CBS limits,
\begin{align}
\label{CBS_3}
\Delta V_{123}^\mathrm{core/CBS} ={}& \frac{3^3 \Delta V_{123}^{\mathrm{core}/X=3}-2^3 \Delta V_{123}^{\mathrm{core}/X=2}}{3^3-2^3},\\
\Delta V_{123}^\mathrm{rel/CBS} ={}& \frac{4^3 \Delta V_{123}^{\mathrm{rel}/X=4}-3^3 \Delta V_{123}^{\mathrm{rel}/X=3}}{4^3-3^3}.
\end{align}
For most configurations, the absolute values of $\Delta V_{123}^\mathrm{core}$ and $\Delta V_{123}^\mathrm{rel}$ are much smaller than those of $\Delta V_{123}^\mathrm{T-(T)}$ and $\Delta V_{123}^\mathrm{(Q)-T}$.

To give an impression of the importance of each of the six contributions to $\Delta V_{123}$, Table~\ref{tab:dV3example}
\begin{table*}
\footnotesize
\caption{\label{tab:dV3example} Contributions to the nonadditive three-body interaction energy $\Delta V_{123}$ (in kelvin) for selected equilateral triangular configurations.} 
\begin{ruledtabular}
\begin{tabular}{d{1.1}d{5.5}d{3.5}d{1.5}d{1.5}d{2.5}d{2.5}d{5.5}}
\multicolumn{1}{c}{$R_{12}/\mathrm{\AA}$} &
\multicolumn{1}{c}{$\Delta V_{123}^\mathrm{SCF}$} &
\multicolumn{1}{c}{$\Delta V_{123}^\mathrm{CCSD(T)c}$} &
\multicolumn{1}{c}{$\Delta V_{123}^\mathrm{T-(T)}$} &
\multicolumn{1}{c}{$\Delta V_{123}^\mathrm{(Q)-T}$} &
\multicolumn{1}{c}{$\Delta V_{123}^\mathrm{core}$} &
\multicolumn{1}{c}{$\Delta V_{123}^\mathrm{rel}$} &
\multicolumn{1}{c}{$\Delta V_{123}$ (total)}
\\[0.5ex]
\hline
\\[-2ex]
1.6 & -4556.92547 & 232.28655 & 7.49920 & 2.69892 & 9.18490  & 23.43675 & -4281.81916 \\
2.0 & -396.34798  & 64.37436  & 2.61186 & 1.12874 & 0.77163  & 2.22235  & -325.23903  \\
2.4 & -31.71760   & 14.42730  & 0.63842 & 0.27579 & 0.02300  & 0.18368  & -16.16940   \\
2.8 & -2.32745    & 3.01487   & 0.14620 & 0.05692 & -0.01010 & 0.01571  & 0.89614     \\
3.2 & -0.16004    & 0.68935   & 0.03883 & 0.01384 & -0.00400 & 0.00206  & 0.58004     \\
3.6 & -0.01056    & 0.19463   & 0.01261 & 0.00439 & -0.00132 & 0.00050  & 0.20025     \\
4.0 & -0.00068    & 0.06815   & 0.00479 & 0.00167 & -0.00047 & 0.00017  & 0.07363     \\
4.4 & -0.00004    & 0.02776   & 0.00203 & 0.00071 & -0.00019 & 0.00007  & 0.03034     \\
4.8 & 0.00000     & 0.01247   & 0.00093 & 0.00032 & -0.00009 & 0.00003  & 0.01366     \\
5.2 & 0.00000     & 0.00600   & 0.00046 & 0.00016 & -0.00004 & 0.00002  & 0.00658     \\
6.0 & 0.00000     & 0.00163   & 0.00013 & 0.00004 & -0.00001 & 0.00000  & 0.00179
\end{tabular}
\end{ruledtabular}
\end{table*}
lists them for selected equilateral triangular configurations. Detailed numerical results for all 2748 considered configurations are provided in the \textcolor{blue}{supplementary material}.


\subsection{Uncertainty budget}
\label{sec:2.2}


The combined uncertainties of the \emph{ab initio} calculated nonadditive three-body interaction energies $\Delta V_{123}$ were determined as the square roots of the sums of the squared uncertainties resulting from the individual contributions. This is in accordance with the standard procedure for the analysis of measurement uncertainties.

The uncertainties of the SCF contribution $\Delta V_{123}^\mathrm{SCF}$ are very small and were neglected. For $\Delta V_{123}^\mathrm{CCSD(T)c}$, the uncertainties were estimated as the absolute values of the differences between the CBS extrapolated values and the non-extrapolated ones, i.e., those obtained at the $X=6$ basis set level. For the contributions $\Delta V_{123}^\mathrm{T-(T)}$, $\Delta V_{123}^\mathrm{(Q)-T}$, and $\Delta V_{123}^\mathrm{core}$, which were obtained with much smaller basis sets, we used twice the absolute values of the differences between the extrapolated and non-extrapolated values. The resulting uncertainties for $\Delta V_{123}^\mathrm{(Q)-T}$ should be conservative enough to also account for the neglect of post-\mbox{CCSDT(Q)} contributions. For $\Delta V_{123}^\mathrm{rel}$, the differences between the extrapolated and non-extrapolated values are, even when doubled, too small to be used as estimates for the uncertainty, since some potentially relevant relativistic effects (such as the orbit-orbit interaction) are not accounted for by the DPT2 approach. Therefore, the uncertainties of $\Delta V_{123}^\mathrm{rel}$ were estimated as twice the absolute values of the differences between the extrapolated values and those obtained with the smaller of the two basis sets. The resulting combined uncertainties of the total $\Delta V_{123}$ values are listed in the \textcolor{blue}{supplementary material}.

We regard all uncertainties as standard uncertainties. Thus, they correspond to a coverage factor of $k=1$ with a confidence level of approximately 68\%.


\subsection{Analytical potential function}
\label{sec:2.3}


As discussed above, 198 of the 2748 \emph{ab initio} values for $\Delta V_{123}$ were discarded due to numerical issues, leaving 2550 $\Delta V_{123}$ values as the basis for the fit of an analytical potential function. The chosen functional form was partly inspired by expressions previously used in the  literature\cite{Lotrich:1997,Schwerdtfeger:2009} and is given as
\begin{align}
\label{eq:Pot3}
\Delta V_{123} ={}& \frac{\nu}{R_\mathrm{g}^9} \left[1+3\cos\theta_1\cos\theta_2\cos\theta_3\right] f_3\left(R_{12}\right)f_3\left(R_{13}\right)f_3\left(R_{23}\right)\nonumber\\
& + \sum_{0\leqslant l_1\leqslant l_2\leqslant l_3}\left(A_{l_1l_2l_3}+B_{l_1l_2l_3}R_\mathrm{g}^2+C_{l_1l_2l_3}R_\mathrm{g}^4+D_{l_1l_2l_3}R_\mathrm{g}^6\right)\nonumber\\
& \times \exp\left[-\alpha_{l_1l_2l_3}\left(R_{12}+R_{13}+R_{23}\right)\right]\nonumber\\
& \times \sum_{(k_1,k_2,k_3)} \left[P_{k_1}\left(\cos\theta_1\right) P_{k_2}\left(\cos\theta_2\right) P_{k_3}\left(\cos\theta_3\right)\right],
\end{align}
where $\nu=11.92$\;a.u.\cite{Kumar:2010} is the triple-dipole dispersion coefficient of neon, $R_\mathrm{g}=\left(R_{12}R_{13}R_{23}\right)^{1/3}$, $f_3(R_{ij})=1-\exp\left(-bR_{ij}\right)\sum_{k=0}^{3}\left(bR_{ij}\right)^k/k!$ is a Tang--Toennies damping function,\cite{Tang:1984} $P_k(x)$ is a $k$th-order Legendre polynomial, the triples ($k_1,k_2,k_3$) are the distinguishable permutations of the values of $l_1$, $l_2$, and $l_3$, and $A_{l_1l_2l_3}$, $B_{l_1l_2l_3}$, $C_{l_1l_2l_3}$, $D_{l_1l_2l_3}$, $\alpha_{l_1l_2l_3}$, and $b$ are adjustable parameters. At large interatomic separations, the analytical potential function approaches the Axilrod--Teller--Muto (ATM) potential,\cite{Axilrod:1943,Muto:1943}
\begin{equation}
\label{eq:ATM}
\Delta V_{123}^\mathrm{ATM} = \frac{\nu}{R_\mathrm{g}^9}\left[1+3\cos\theta_1\cos\theta_2\cos\theta_3\right],
\end{equation}
which represents the physically correct asymptotic behavior of the nonadditive three-body interactions between three atoms.

We restricted the sum over $l_1$, $l_2$, and $l_3$ in Eq.~\eqref{eq:Pot3} such that $l_1+l_2+l_3 \leqslant 7$, and we excluded the case $l_1=l_2=l_3=1$ because the resulting product of three first-order Legendre polynomials can be expressed as a linear combination of zeroth- and second-order Legendre polynomials. With these restrictions, there are 30 combinations of $l_1$, $l_2$, and $l_3$ values and thus a total of 151 adjustable parameters, which were fully optimized in a non-linear least-squares fit to the 2550 \emph{ab initio} points. To ensure a balanced fit quality over the entire potential energy surface, each point was weighted by
\begin{equation}
\label{eq:w}
\frac{\left(R_\mathrm{g}/\mathrm{\AA}\right)^{9}}{\left[1+10^{-6}\left(V_{123}/\mathrm{K}+126\right)^2\right]^2},
\end{equation}
where we note that the maximum depth of the total three-body interaction potential $V_{123}$ (which is the sum of the three pair potentials and the nonadditive three-body contribution) is approximately $126$\;K, so that the denominator becomes smallest at the minimum of $V_{123}$. We tested for possible overfitting by also performing a fit of the same function to a vastly reduced number of points obtained by discarding 72 of the 156 triangular shapes, namely, every other group of 12 shapes for which in Fig.~\ref{fig:triangles} the positions of atom 3 lie on a straight line. The discarded points were still described almost as well as the points retained in the fit. Furthermore, we visually inspected many one-dimensional cuts through the final potential function and found it to be very smooth with no apparent unphysical ``wiggles,'' whose appearance would have indicated overfitting.

\begin{figure*}
\includegraphics{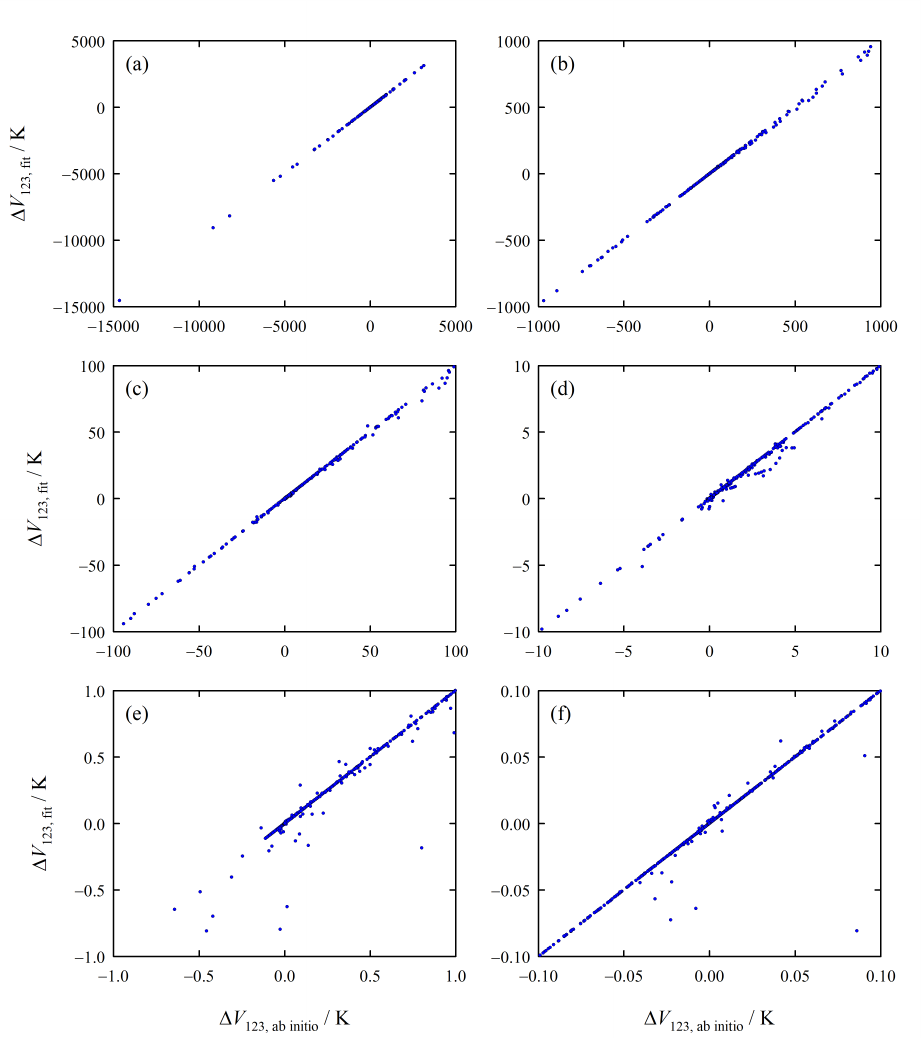}
\caption{Nonadditive three-body interaction energies obtained with the fitted potential function versus the respective \emph{ab initio} calculated values for different energy ranges [(a) to (f)].}
\label{fig:fit}
\end{figure*}

In Fig.~\ref{fig:fit},
the $\Delta V_{123}$ values obtained with the analytical potential function are plotted against the corresponding \emph{ab initio} calculated values for different energy ranges. The figure shows that most points lie very close to the diagonal. Those points that deviate more markedly fall into two categories, and in both cases the deviations are no cause for concern. One category of points are those for which the absolute value of $\Delta V_{123}$ is very small because the potential changes sign very close to the point, so that large relative deviations (and thus large deviations from the diagonal) are to be expected. The other category of points are those for which $R_{12}$ is so small that atoms 1 and 2 strongly repel each other, while $R_{13}$ and $R_{23}$ are much larger. These points, rightfully, had a very low weight in the fit due to the large value of $V_{12}$ (and thus of the total potential energy $V_{123}$), which completely dominates the interaction between the three atoms. For example, for the configuration with $R_{12}=1.6$\;\AA{}, $R_{13}=3.68$\;\AA{}, and $\theta_1=168^\circ$, the \emph{ab initio} value for $\Delta V_{123}$ is $0.08604$\;K, whereas the value obtained with the fitted potential function is $-0.08087$\;K, see the point in the lower right corner of Fig.~\ref{fig:fit}(f). However, when we take into account that the pair interaction energies $V_{12}$, $V_{13}$, and $V_{23}$ have values of $26851.17$\;K, $-22.29$\;K, and $-2.44$\;K, respectively, it is clear that the failure of the analytical function to reproduce the \emph{ab initio} value for $\Delta V_{123}$ for this configuration is unproblematic.

Because the effects of positive and negative fit errors on thermophysical properties such as virial coefficients are expected to largely cancel, we decided to not include the fit errors in our uncertainty budget for $\Delta V_{123}$. To propagate the uncertainties of $\Delta V_{123}$ into uncertainties for the virial coefficients, we fitted the analytical function also to the \emph{ab initio} values for $\Delta V_{123}$ shifted by either adding or subtracting their combined standard uncertainties. The resulting $\Delta V_{123}$ values obtained with these two additional functions are denoted as $\Delta V_{123}^+$ and $\Delta V_{123}^-$. In the \textcolor{blue}{supplementary material}, we provide a Fortran~90 code for evaluating $\Delta V_{123}$, $\Delta V_{123}^+$, and $\Delta V_{123}^-$.

\begin{figure*}
\includegraphics{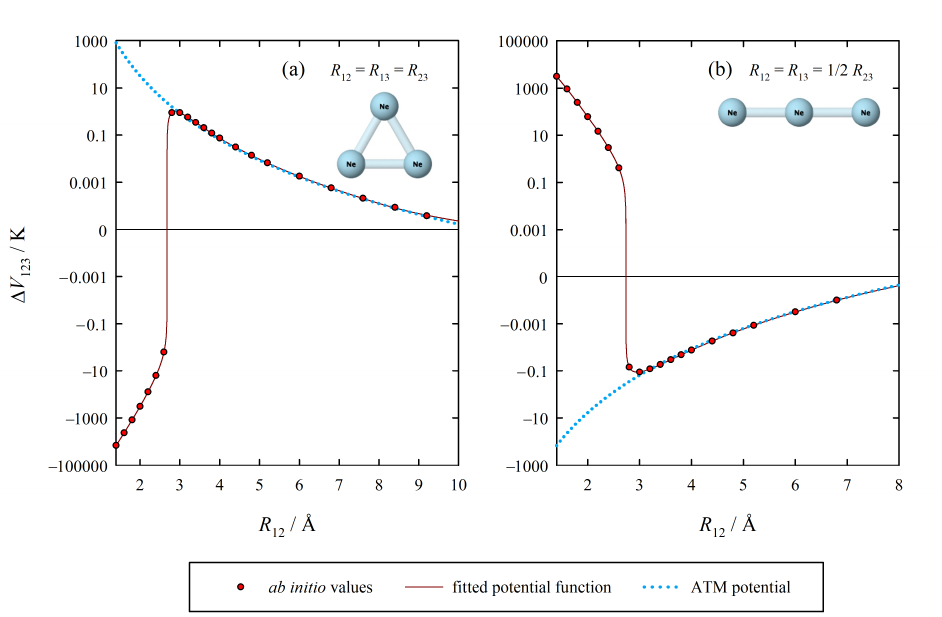}
\caption{Nonadditive three-body interaction energies obtained from the quantum-chemical \emph{ab initio} calculations, from the fitted potential function, and from the simple ATM potential for two of the 156 investigated triangular shapes: (a) equilateral triangles, (b) symmetric linear configurations.}
\label{fig:dV123}
\end{figure*}

Figure~\ref{fig:dV123}
shows $\Delta V_{123}$ and $\Delta V_{123}^\mathrm{ATM}$ as a function of $R_{12}$ for two of the 156 investigated triangular shapes, namely, equilateral triangles and symmetric linear configurations. While in both cases $\Delta V_{123}^\mathrm{ATM}$ is a good approximation of the true nonadditive three-body potential for $R_{12}$ values larger than approximately 3\;\AA{} (which corresponds closely to the minimum of $V_{12}$), the short-range behavior of $\Delta V_{123}^\mathrm{ATM}$ is clearly incorrect as it does not account for nonadditive exchange interactions.


\section{Nonadditive four-body potential}
\label{sec:3}



\subsection{\emph{Ab initio} calculations}
\label{sec:3.1}


The nonadditive four-body potential of four neon atoms is a six-dimensional surface. An adequate sampling of such a high-dimensional surface by quantum-chemical \emph{ab initio} calculations would probably require tens of thousands of grid points. Because nonadditive four-body interactions can be expected to contribute very little to the fourth density and acoustic virial coefficients, we decided to treat these interactions in a highly simplified manner by only investigating configurations in which the four atoms form a regular tetrahedron. We considered a total of 40 edge lengths in the range from 1.4 to 4.5\;\AA{}. The individual nonadditive four-body interaction energies were obtained using again the supermolecular approach,
\begin{align}
\label{eq:supermolecule4}
\Delta V_{1234} ={}& V_{1234} - V_{12} - V_{13} - V_{14} - V_{23} - V_{24} - V_{34} \nonumber\\
& - \Delta V_{123} - \Delta V_{124} - \Delta V_{134} - \Delta V_{234} \nonumber\\
={}& E_{1234} - E_{123} - E_{124} - E_{134} - E_{234} \nonumber\\ 
& + E_{12} + E_{13} + E_{14} + E_{23} + E_{24} + E_{34} \nonumber\\
& - E_{1} - E_{2} - E_{3} - E_{4},
\end{align}
where the electronic energies were calculated with CFOUR\cite{cfour,Matthews:2020} at the \mbox{FC-CCSD(T)} level of theory using the \mbox{aug-cc-pV$X$Z} basis sets up to \mbox{$X=5$} for the neon atoms and an additional ``ghost'' hydrogen atom placed in the center of the tetrahedron to improve the convergence towards the CBS limit.\cite{Patkowski:2012} The counterpoise procedure\cite{Boys:1970} was applied to avoid the BSSE, i.e., all energies were obtained in the full basis set of the four-atom system with the additional hydrogen basis functions in the center. The resulting $\Delta V_{1234}$ values for $X=4$ and $X=5$ turned out to be very close, partly even within the numerical noise. Therefore, we did not perform a CBS extrapolation and used the $X=5$ results as our best estimates of $\Delta V_{1234}$. The calculated $\Delta V_{1234}$ values for all basis set levels are provided in the \textcolor{blue}{supplementary material}.


\subsection{Analytical potential function}
\label{sec:3.2}


The starting point for our analytical fit of the nonadditive four-body interaction energies is the Bade potential,\cite{Bade:1958} which is the nonadditive four-body analog of the ATM nonadditive three-body potential and is given as
\begin{equation}
\label{eq:Bade}
\Delta V_{1234}^\mathrm{Bade} = \omega \left(f_{1234}+f_{1243}+f_{1324}\right),
\end{equation}
with
\begin{align}
\label{eq:fijkl}
f_{ijkl} ={}& \frac{1}{\left(R_{ij}R_{jk}R_{kl}R_{il}\right)^{3}}\big[ \left(\mathbf{\hat{u}}_{ij}\cdot\mathbf{\hat{u}}_{jk}\right)^2 + \left(\mathbf{\hat{u}}_{ij}\cdot\mathbf{\hat{u}}_{kl}\right)^2 + \left(\mathbf{\hat{u}}_{ij}\cdot\mathbf{\hat{u}}_{il}\right)^2  \nonumber\\
& + \left(\mathbf{\hat{u}}_{jk}\cdot\mathbf{\hat{u}}_{kl}\right)^2 + \left(\mathbf{\hat{u}}_{il}\cdot\mathbf{\hat{u}}_{jk}\right)^2 + \left(\mathbf{\hat{u}}_{il}\cdot\mathbf{\hat{u}}_{kl}\right)^2  \nonumber\\
& - 3\left(\mathbf{\hat{u}}_{ij}\cdot\mathbf{\hat{u}}_{jk}\right)\left(\mathbf{\hat{u}}_{jk}\cdot\mathbf{\hat{u}}_{kl}\right)\left(\mathbf{\hat{u}}_{ij}\cdot\mathbf{\hat{u}}_{kl}\right)  \nonumber\\
& - 3\left(\mathbf{\hat{u}}_{ij}\cdot\mathbf{\hat{u}}_{jk}\right)\left(\mathbf{\hat{u}}_{il}\cdot\mathbf{\hat{u}}_{jk}\right)\left(\mathbf{\hat{u}}_{ij}\cdot\mathbf{\hat{u}}_{il}\right)  \nonumber\\
& - 3\left(\mathbf{\hat{u}}_{ij}\cdot\mathbf{\hat{u}}_{kl}\right)\left(\mathbf{\hat{u}}_{il}\cdot\mathbf{\hat{u}}_{kl}\right)\left(\mathbf{\hat{u}}_{ij}\cdot\mathbf{\hat{u}}_{il}\right) \nonumber\\
& - 3\left(\mathbf{\hat{u}}_{jk}\cdot\mathbf{\hat{u}}_{kl}\right)\left(\mathbf{\hat{u}}_{il}\cdot\mathbf{\hat{u}}_{kl}\right)\left(\mathbf{\hat{u}}_{il}\cdot\mathbf{\hat{u}}_{jk}\right)  \nonumber\\
& + 9\left(\mathbf{\hat{u}}_{ij}\cdot\mathbf{\hat{u}}_{jk}\right)\left(\mathbf{\hat{u}}_{jk}\cdot\mathbf{\hat{u}}_{kl}\right)\left(\mathbf{\hat{u}}_{il}\cdot\mathbf{\hat{u}}_{kl}\right)\left(\mathbf{\hat{u}}_{ij}\cdot\mathbf{\hat{u}}_{il}\right) - 1\big],
\end{align}
where $\omega$ is the quadruple-dipole dispersion coefficient and $\mathbf{\hat{u}}_{ij}$ is the unit vector pointing from atom $i$ towards atom $j$.

The Bade potential correctly describes the long-range asymptotic behavior of the nonadditive four-body dispersion interactions, but, just like the ATM potential, it does not account for exchange interactions and thus has the wrong short-range behavior. To remedy this issue, following Schwerdtfeger \emph{et~al.},\cite{Schwerdtfeger:2017} we fitted a modification of the Bade potential with additional parameters for increased flexibility to the 40 \emph{ab initio} calculated values for $\Delta V_{1234}$. This extended Bade potential, which retains the correct long-range asymptotic behavior of the simple Bade potential, is given by
\begin{equation}
\label{eq:eBade}
\Delta V_{1234} = g_{1234} \Delta V_{1234}^\mathrm{Bade},
\end{equation}
with
\begin{equation}
\label{eq:g1234}
g_{1234} = 1+\exp\left(-\delta R_\mathrm{a}\right)\sum_{k=0}^{6}A_kR_\mathrm{g}^k,
\end{equation}
where $R_\mathrm{a}=\left(R_{12}+R_{13}+R_{14}+R_{23}+R_{24}+R_{34}\right)/6$, $R_\mathrm{g}=\left(R_{12}R_{13}R_{14}R_{23}R_{24}R_{34}\right)^{1/6}$, and $\delta$ and $A_0$ to $A_6$ are fit parameters. Since $\omega$ for neon is not precisely known, we treated it as a fit parameter as well. Note that the function $g_{1234}$ as given by Eq.~\eqref{eq:g1234} differs slightly from that used by Schwerdtfeger \emph{et~al.}\cite{Schwerdtfeger:2017} A Fortran~90 code for evaluating $\Delta V_{1234}$ is provided in the \textcolor{blue}{supplementary material}.

\begin{figure}
\includegraphics{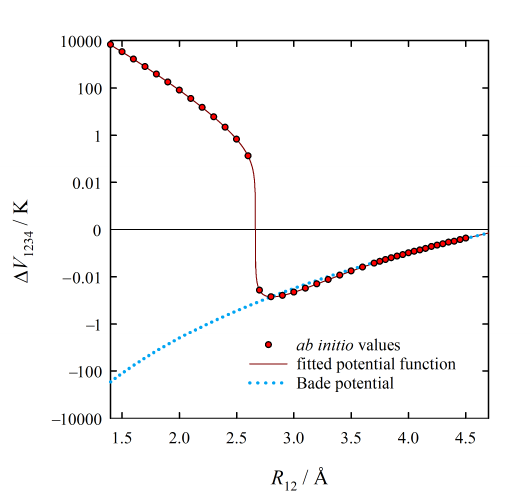}
\caption{Nonadditive four-body interaction energies for regular tetrahedra of neon atoms obtained from the quantum-chemical \emph{ab initio} calculations, from the fitted extended Bade potential function, and from the simple Bade potential.}
\label{fig:dV1234}
\end{figure}

Figure~\ref{fig:dV1234}
depicts $\Delta V_{1234}$ and $\Delta V_{1234}^\mathrm{Bade}$ for the investigated regular tetrahedra of neon atoms. Similar to the case for $\Delta V_{123}$ and $\Delta V_{123}^\mathrm{ATM}$ (see Fig.~\ref{fig:dV123}), $\Delta V_{1234}$ and $\Delta V_{1234}^\mathrm{Bade}$ behave very differently at short interatomic separations.


\section{Calculation of virial coefficients}
\label{sec:4}
  The virial coefficients were calculated at temperatures from 10 to 5000\;K using the path-integral Monte Carlo (PIMC) approach. A detailed description of this method can
  be found in the original publications;\cite{Jordan:1968,Garberoglio:2009,Garberoglio:2011} here, we
  provide only a brief outline. The PIMC approach exploits the well-known quantum-classical isomorphism of
  quantum statistical mechanics, whereby the wave function of a quantum particle $i$ is mapped onto a
  ring polymer with $P$ monomers. 
  We will denote the coordinates of this ring polymer as $\mbr_p^{(i)}$
  with $p \in [1,P]$ and the understanding that $\mbr_1^{(i)} = \mbr_{P+1}^{(i)} = \mathbf{0}$,
  where $\mathbf{0}$ is the zero vector. This mapping is exact in the $P \to \infty$ limit, but
  well-converged calculations are obtained with a finite value of $P$. In all of our calculations,
  we used a temperature-dependent value of $P$ according to $P = \mathrm{int}\left(4 +
  800/T\right)$, where $\mathrm{int}(x)$ denotes the smallest integer larger than $x$. This choice
  enabled us to converge the PIMC calculations well within the uncertainties propagated from the
  potentials, as discussed below.
  
  The interaction between ring polymers is obtained by averaging
  over the monomers with the same index $p$, and we define
  \begin{equation}
    \overline{V}_{ij} = \frac{1}{P} \sum_{p=1}^P V_{ij}\left(\left|
    \mbR_i + \mbr_p^{(i)} - \mbR_j - \mbr_p^{(j)}
    \right|\right),
    \label{eq:overline_V}
  \end{equation}
  with analogous definitions for the three and four-body potentials $V_{ijk}$ and $V_{ijkl}$.
  
  The quantum-classical isomorphism specifies the distribution of the variables $\mbr_p^{(i)}$ (a
  Brownian process with fixed endpoints, also known as Brownian bridge),
  which can be sampled efficiently with interpolation
  formulae.~\cite{Levy:1954,Fosdick:1966,Ceperley:1995} Unlike the original publications,
  we found it convenient to use atomic positions instead of Jacobi coordinates to sample Monte Carlo
  configurations.\cite{Garberoglio:2011}
  In principle, one should take into account the fermionic or bosonic nature of neon atoms. Although
  these effects can be incorporated into PIMC calculations,~\cite{Garberoglio:2011b} the
  temperatures studied in this work are so high that limiting calculations to Boltzmann statistics
  (distinguishable particles) did not introduce any appreciable difference.
  


\subsection{Density virial coefficients} \label{sec:4.1}
  Defining
  \begin{equation}
    z_2(ij) = \exp\left(-\beta \overline{V}_{ij}\right),
    \label{eq:zk}
  \end{equation}
  where $\beta=(k_\mathrm{B}T)^{-1}$, with analogous definitions for $z_3(ijk)$ and $z_4(ijkl)$, the path-integral expression for $C(T)$
  can be written as
  \begin{equation}
    C(T) = - \frac{N^2_\mathrm{A}}{3} \int \left[
    \left\langle z_3 - 3 z_2 + 2 \right\rangle_3
    - 3 \left\langle (z_2-1)^2 \right\rangle_4 \right] ~ \dd^3\mbR_2 \dd^3\mbR_3,
    \label{eq:C}
  \end{equation}
  where $N_\mathrm{A}$ is Avogadro's constant, and the first particle is fixed at the origin of the
  coordinate system (i.e., $\mbR_1 = \mathbf{0}$). Angular brackets denote the average over
  ring-polymer configurations. In the first term, there is one ring polymer associated with each of
  the three particles. The second term is actually \mbox{$[z_2(12)-1] [z_2(34)-1]$} with $\mbR_4 =
  \mathbf{0}$.

  We finally note that to improve the efficiency of the calculations it is convenient to write $3
  z_2 = z_2(12) + z_2(13) + z_2(23)$ and use rotational invariance to reduce the
  multidimensional integral to a three-dimensional one, $\dd^3\mbR_2 \dd^3\mbR_3 = 8 \pi^2 R_2^2
  R_3^2 \dd R_2 \dd R_3 \dd\cos\theta_{1}$.

  In our calculations of $C(T)$, the three-dimensional integral has been evaluated using a modified parallel
  version~\cite{parallel_vegas:2025} of the GSL~\cite{GSL:2009} VEGAS~\cite{Lepage:1978} routine
  using $2 \times 10^6$ Monte Carlo evaluations and six independent ring-polymer configurations for
  each particle to evaluate the bracket averages in Eq.~(\ref{eq:C}). In this way, the statistical
  uncertainty on $C(T)$ was reduced to less than 5\% of the uncertainty propagated from the
  potentials.
  We computed the cross
    coefficients considering all ten possible isotopic combinations, which where then combined
    to provide the mixture coefficients through the exact relation
    \begin{equation}
      C_\mathrm{mix}(T) = \sum_{i=20}^{22}\sum_{j=20}^{22}\sum_{k=20}^{22} x_i x_j x_k C_{i,j,k}(T),
      \label{eq:averaging}
    \end{equation}
    using the mole fractions of the natural abundances $x_{20}
    = 0.9048$, $x_{21}=0.0027,$ and $x_{22}=0.0925$. The same abundances were used to compute the
    average molar mass $M$, which was in turn used to compute density and acoustic virial
  coefficients using fictitious atoms of mass $M$. The absolute value of the difference between
  the mixture virial coefficients and those computed with the average molar mass divided by the
  uncertainty propagated from the potentials was found to be on average $\approx 4\%$.
   
  The expression of the fourth density virial coefficient is
  \begin{eqnarray}
    D(T) &=& -\frac{N_\mathrm{A}^3}{8} \int \left[
    \left\langle z_4 - 4 z_3 - 3 z_2^2 + 12 z_2 - 6
    \right\rangle_4
    \right. \nonumber \\
    & & 
    -12 \left\langle
    (z_2-1)(z_3 - 3z_2 + 2)
    \right\rangle_5 \nonumber \\
    & & \left.
    +20 \left\langle
    (z_2-1)^3
    \right\rangle_6
    \right] ~ \dd^3\mbR_2 \dd^3\mbR_3 \dd^3\mbR_4,
    \label{eq:D}
  \end{eqnarray}
  where, as in Eq.~(\ref{eq:C}), the first particle is fixed at the origin of the coordinate system,
  and in the average over five and six ring polymers the additional ones are also fixed at the origin of
  the coordinate system. In evaluating $D(T)$, we found it numerically convenient to write it as
  \begin{equation}
    D(T) = D_2(T) + D_{32}(T) + D_{43}(T),
    \label{eq:D_components}
  \end{equation}
  where $D_2(T)$ corresponds to Eq.~(\ref{eq:D}) evaluated considering only the pair potential,
  $D_{32}(T)$ is the nonadditive three-body contribution [i.e., the difference between $D(T)$
    evaluated using only the pair and nonadditive three-body potentials and $D_2(T)$], and
  $D_{43}(T)$ is the nonadditive four-body contribution.

  In our calculations, $D_2(T)$ was evaluated as the average of $16$ independent calculations using
  $2 \times 10^6$ Monte Carlo calls to evaluate the integral and 16 independent ring-polymer
  configurations to evaluate the angular brackets in Eq.~(\ref{eq:D}).  $D_{32}(T)$ was evaluated
  using a single calculation of $10^6$ Monte Carlo steps and 16 independent ring polymers to
  evaluate the angular brackets and, finally, the very small contribution $D_{43}(T)$ was evaluated
    using a single calculation 
  of $10^5$ Monte Carlo steps and eight independent ring polymers.  In this way, the statistical
  uncertainty of $D(T)$ was reduced to less than 5\% of the uncertainty propagated from the
  potentials. Since the calculations of the third density and acoustic virial coefficients
    showed that using the average molar mass provides an excellent approximation to the values for
    the natural isotopic mixture, the calculations of $D(T)$ for natural neon 
    were carried out only within this approximation. In addition, we calculated $D(T)$ for the pure ${}^{20}$Ne isotope.
    
\subsection{Acoustic virial coefficients}
\label{sec:4.2}


Gillis and Moldover\cite{Gillis:1996} derived the expressions for the second, third, and fourth acoustic virial coefficients in terms of the second, third, and fourth density virial coefficients and their first and second temperature derivatives:
\begin{equation}
\label{eq:beta_a}
\beta_\mathrm{a} = 2B + 2\left(\gamma^\circ-1\right)B_\mathrm{t} + \frac{\left(\gamma^\circ-1\right)^2}{\gamma^\circ}B_\mathrm{tt},
\end{equation}
\begin{align}
\label{eq:gamma_a}
RT\gamma_\mathrm{a} ={}& \frac{\gamma^\circ-1}{\gamma^\circ}Q^2-\beta_\mathrm{a}B + \frac{2\gamma^\circ+1}{\gamma^\circ}C + \frac{\gamma^{\circ 2}-1}{\gamma^\circ}C_\mathrm{t} \nonumber\\
& + \frac{\left(\gamma^\circ-1\right)^2}{2\gamma^\circ}C_\mathrm{tt},
\end{align}
\begin{align}
\label{eq:delta_a}
\left(RT\right)^2\delta_\mathrm{a} ={}& \frac{\left(\gamma^\circ-1\right)^2}{\gamma^\circ}Q^2\left(2B_\mathrm{t}+B_\mathrm{tt}\right) + \frac{\gamma^\circ-1}{\gamma^\circ}QP - \beta_\mathrm{a}C \nonumber\\
& - 2RT\gamma_\mathrm{a}B + \frac{2\left(\gamma^\circ+1\right)}{\gamma^\circ}D + \frac{2\left(\gamma^\circ-1\right)\left(\gamma^\circ+2\right)}{3\gamma^\circ}D_\mathrm{t} \nonumber\\
& + \frac{\left(\gamma^\circ-1\right)^2}{3\gamma^\circ}D_\mathrm{tt},
\end{align}
where
\begin{align}
\label{eq:Q}
Q ={}& B + \left(2\gamma^\circ-1\right)B_\mathrm{t} + \left(\gamma^\circ-1\right) B_\mathrm{tt}, \\
P ={}& 2C + 2\gamma^\circ C_\mathrm{t} + \left(\gamma^\circ-1\right) C_\mathrm{tt},
\end{align}
and
\begin{equation}
\label{eq:Xt_Xtt}
X_\mathrm{t}=T\frac{\mathrm{d}X}{\mathrm{d}T},\qquad X_\mathrm{tt}=T^2\frac{\mathrm{d}^2X}{\mathrm{d}T^2}.
\end{equation}
Gillis and Moldover also provided an expression for the fifth acoustic virial coefficient $\left(RT\right)^3\epsilon_\mathrm{a}$, which we mention here because we discovered an error in their expression. The first term in their Eq.~(10) reads $\left(\gamma^\circ-1\right)^3Q^2\left(2B_\mathrm{t}+B_\mathrm{tt}\right)^2B_\mathrm{tt}$, but it should read $\left(\gamma^\circ-1\right)^3Q^2\left(2B_\mathrm{t}+B_\mathrm{tt}\right)^2$. We checked that the corrected expression yields the same values for $\left(RT\right)^3\epsilon_\mathrm{a}$ as an alternative, less compact expression for $\left(RT\right)^3\epsilon_\mathrm{a}$ derived by Gokul \emph{et~al.}\cite{Gokul:2021}

  The path-integral expressions for the acoustic virial coefficients can be derived from the
  definitions above using Eqs.~(\ref{eq:C}), (\ref{eq:D}), and the path-integral expression for
  $B(T)$,
  \begin{equation}
    B(T) = -\frac{N_\mathrm{A}}{2} \int \left\langle z_2 - 1 \right\rangle ~ \dd^3 \mbR_2.
    \label{eq:B}
  \end{equation}
  In performing the temperature derivatives, one has to consider the fact that the probability
  distribution for the ring-polymer configurations is temperature dependent. This leads to very long
  and cumbersome equations that, to the best of our knowledge, have been derived only in the case of
  $RT\gamma_\mathrm{a}$, and we refer the reader to the original
  papers.~\cite{Garberoglio:2023,Lang:2023} An additional complication in deriving path-integral
  expressions for the acoustic virial coefficients comes from the fact that straightforward
  differentiation with respect to the temperature leads to expressions mathematically similar to the
  thermodynamic estimator of the kinetic energy in the path-integral formalism, which is known to
  have a large variance. It is convenient to further express the resulting equations in a form
  similar to the virial estimator of the kinetic energy, which enables the calculation of the
  acoustic virial coefficients with considerably less computational resources.~\cite{Lang:2023}  

  For these reasons, we performed path-integral calculations only for $RT\gamma_\mathrm{a}(T)$ in this
  work. Analogous to the case of $D(T)$, also in this case it is computationally convenient to
  express the third acoustic virial coefficient as the sum of a pair contribution and a remainder
  depending on the non-additive three-body potential. The two-body contribution to
  $RT\gamma_\mathrm{a}(T)$ has been computed by averaging $96$ independent runs, each using $10^6$
  Monte Carlo steps and 16 independent ring polymers to evaluate the angular brackets. The
  nonadditive three-body contribution was evaluated by averaging $48$ independent runs of $10^5$
  Monte Carlo steps each and using eight independent ring polymers to evaluate the angular
  brackets. In this case, the ratio between the statistical uncertainty of the PIMC
    calculation and the uncertainty propagated from the potentials was found to be $\approx 15\%$ on average.

  Analogously to what has been done in the case of $C(T)$, we computed $RT\gamma_\mathrm{a}(T)$ for
  all ten possible isotopic combinations and for the average molar mass $M$. In this case, the
  absolute value of the difference between the mixture virial coefficients and those computed with
  the average molar mass divided by the uncertainty propagated from the potentials was found to be
  on average $\approx 15\%$.
  
\subsection{Uncertainty propagation} \label{sec:4.3}

  The traditional method~\cite{Garberoglio:2009} to propagate the uncertainty from the (non-additive) potentials to the
  uncertainty of the virial coefficient has been to compute the virial coefficients with rigidly
  shifted potentials and estimate
  \begin{equation}
    u_N(X) = \frac{1}{2} \left| X_+(T) - X_-(T) \right| \approx
    \left | \int \delta v_N \frac{\delta X}{\delta v_N} ~ d\Omega \right|, 
    \label{eq:unc_propagation}
  \end{equation}
  where $u_N(X)$ denotes the standard ($k=1$) uncertainty on the density virial coefficient $X$ ($X
  = B, C, D, \ldots$) coming from the $N$-body (non-additive) potential $v_N$, and $\delta v_N$ is
  the standard uncertainty of the $N$-body potential (which is a positive quantity).  The last term
  in Eq.~(\ref{eq:unc_propagation}), which becomes an equality in the limit $\delta v_N \ll |v_N|$,
  shows that the uncertainty can also be interpreted as a functional variation with respect to the
  $N$-body potential; we denoted by $\Omega$ all the variables entering the integrals leading to
  $X$.  The total uncertainty of the virial coefficient $X$ is obtained by summing in quadrature the
  uncertainties $u_N(X)$ for all relevant values of $N$ [i.e., $N=2,3$ for $C(T)$ and $N=2,3,4$ for
    $D(T)$].
  
  However, the approach for uncertainty propagation outlined in Eq.~(\ref{eq:unc_propagation}) is
  unsatisfactory for a few reasons: first of all, it assumes a rigid shift of the potentials for all
  possible particle configurations. However, due to the way that the potential uncertainty is
  estimated, the actual $N$-body potential can be systematically larger than the expected value for
  some configurations and smaller for others. Additionally, the argument of the
    absolute value in Eq.~(\ref{eq:unc_propagation}) can change sign as a function of temperature,
    hence for some specific values of $T$ one would unreasonably be led to conclude that $u_N(X) =
    0$.
  
  One way to overcome these shortcomings is to propagate the uncertainty by taking the absolute
  value of the integrand in the right-hand side of
  Eq.~(\ref{eq:unc_propagation}),~\cite{Garberoglio:2021b,Wheatley:2023} that is
  \begin{equation}
    u_N(X) = \int \delta v_N \left| \frac{\delta X}{\delta v_N} \right| ~ d\Omega.
    \label{eq:unc_propagation_absolute}
  \end{equation}
  This choice is equivalent to assuming that the $N$-body potential is overestimated for those
  configurations where the functional derivative $\delta X / \delta v_N$ is negative (so that one
  should consider a negative variation $\delta v_N$) and underestimated when the
  functional derivative is positive (so that one considers a positive variation $\delta v_N$). In
  this way, $u_N(X)$ turns out to be always nonzero.
  From these considerations, it is clear that Eq.~(\ref{eq:unc_propagation_absolute}) is an upper
  bound to the actual propagated uncertainty, assuming that the actual $N$-body potential is always
  contained within $v_N - \delta v_N$ and $v_N + \delta v_N$. In practice, estimating
    the uncertainty using Eq.~(\ref{eq:unc_propagation_absolute}) provides values that can be much
    larger, especially at low temperatures, than those obtained using Eq.~(\ref{eq:unc_propagation}).
  
  From Eqs.~(\ref{eq:C}), (\ref{eq:D}), and (\ref{eq:B}) one can see that what is needed to compute
  $u_N(X)$ according to Eq.~(\ref{eq:unc_propagation_absolute}) are just
  the variations of the quantities $z_k$ [see Eq.~(\ref{eq:zk})] with respect to the potentials
  $v_N$.  Denoting these variations by $\left. \delta z_k \right|_N$, one has
  \begin{eqnarray}
    \left. \delta z_2(ij) \right|_2 &=& -\beta \delta \overline{V}_{ij}   ~z_2(ij)
    \equiv -\beta \delta V_2 ~ z_2, \\
    \left. \delta z_3(ijk) \right|_2 &=& -\beta \left( \delta \overline{V}_{ij} + \delta
    \overline{V}_{ik} + \delta \overline{V}_{jk} \right) ~z_3 \\
    &\equiv& -3 \beta \delta V_2 ~ z_3, \nonumber \\    
    \left. \delta z_3(ijk) \right|_3 &=& -\beta \delta \Delta \overline{V}_{ijk}   ~z_3
    \equiv -\beta \delta V_3 ~ z_3, \\
    \left. \delta z_4(ijkl) \right|_2 &=& -\beta \left(
    \delta \overline{V}_{ij} + \delta \overline{V}_{ik} + \delta \overline{V}_{il}
    \right. \nonumber \\
    & & \left. 
    + \delta \overline{V}_{jk} + \delta \overline{V}_{jl} + \delta \overline{V}_{kl}    
    \right) ~z_4 \\
    &\equiv&  -6 \beta \delta V_2 ~ z_4, \nonumber \\
    \left. \delta z_4(ijkl) \right|_3 &=& -\beta \left(
    \delta \Delta \overline{V}_{jkl} +
    \delta \Delta \overline{V}_{ikl} \right. \nonumber \\
    & & \left.
    + \delta \Delta \overline{V}_{ijl} +
    \delta \Delta \overline{V}_{jkl}
    \right) z_4 \\
    &\equiv& -4 \beta \delta V_3 ~ z_4, \nonumber \\
    \left. \delta z_4(ijkl) \right|_4 &=& -\beta \delta \overline{V}_{ijkl} ~ z_4
    \equiv -\beta \delta V_4 ~ z_4,
  \end{eqnarray}
  where the equivalence sign ($\equiv$) defines a compact notation. The resulting formulae for the
  propagated uncertainties from the $N$-body non-additive potentials to the second, third, and fourth
  density virial coefficients are
  \begin{eqnarray}
    u_2(B) &=& \frac{\beta N_\mathrm{A}}{2}
    \int \langle \delta V_2 ~ z_2 \rangle ~ \dd^3 \mbR_2, \\
    u_2(C) &=& \frac{\beta N_\mathrm{A}^2}{3}
    \int \left| \langle 3 \delta V_2 ~ ( z_3 - z_2 )\rangle_3  \right.\nonumber \\
    & & - \left. \langle 6 \delta V_2 ~ z_2 (z_2 - 1) \rangle_4 \right| ~\dd^3 \mbR_2 \dd^3
    \mbR_3, \\ 
    u_3(C) &=& \frac{\beta N_\mathrm{A}^2}{3}
    \int \left\langle \delta V_3 ~ z_3 \right\rangle_3 ~ \dd^3 \mbR_2 \dd^3 \mbR_3, \\
    u_2(D) &=& \frac{\beta N_\mathrm{A}^3}{8} \int \left|
    \left\langle 6 \delta V_2 (z_4 - 2 z_3 - z_2^2 + 2 z_2)\right\rangle_4 \right. \nonumber \\
    & &- 12 \left\langle \delta V_2 \left[ z_2 (z_3 - 3z_2 + 2) - 3(z_2-1)(z_3 - z_2)\right] \right\rangle_5
    \nonumber \\
    & &+\left. 60 \left\langle \delta V_2 ~ z_2 (z_2 -1)^2 \right\rangle_6 \right|
    \dd^3 \mbR_2 \dd^3 \mbR_3\dd^3 \mbR_4, \\
    u_3(D) &=& \frac{\beta N_\mathrm{A}^3}{8} \int \left|
    4 \left\langle \delta V_3 (z_4 - z_3) \right\rangle_4 \right. \nonumber \\
    & & -\left. 12 \left\langle \delta V_3 ~ z_3 (z_2-1) \right\rangle_5 \right| ~
    \dd^3 \mbR_2 \dd^3 \mbR_3\dd^3 \mbR_4,\\  
    u_4(D) &=& \frac{\beta N_\mathrm{A}^3}{8}
    \int \left\langle \delta V_4~ z_4 \right\rangle_4 ~
    \dd^3 \mbR_2 \dd^3 \mbR_3\dd^3 \mbR_4,
  \end{eqnarray}
  and hence, for the combined uncertainties from the potentials,
  \begin{eqnarray}
    u(B) &=& u_2(B), \\
    u(C) &=& \sqrt{u_2(C)^2 + u_3(C)^2}, \\
    u(D) &=& \sqrt{u_2(D)^2 + u_3(D)^2 + u_4(D)^2}.
  \end{eqnarray}
  The uncertainty propagation for the third acoustic virial coefficient can be
    computed in an analogous manner. The derivation is somewhat cumbersome and is reported in
    Appendix~\ref{sec:appendix}. 
  
  In principle, one should also consider the statistical uncertainty of the PIMC calculation as
  a contribution to the combined uncertainty.
  However, since the statistical uncertainty in this work has been reduced on average to less than $15\%$ of the
  uncertainty from the potentials, we neglect the statistical contribution.
  While the uncertainty propagated from the potential via path-integral calculations exhibits a
    weak dependence on the isotope mass, the resulting differences are negligible. Consequently, we
    propagated the uncertainty using only the ${}^{20}$Ne isotope.
  
    The uncertainty propagated from the pair potential is the largest contribution
    for all virial coefficients. The relative contribution from the nonadditive three-body potential is found
    to be an increasing function of the temperature. The ratio between these two uncertainties is $\approx
    10\%$ at low temperature in all cases. It increases to $\approx 70\%$ for $C$,
    $\approx 50\%$ for $RT \gamma_\mathrm{a}$, and $\approx 90\%$ for $D$ at the highest
    temperature investigated (5000\;K).
    We have no way to estimate the uncertainty due to the nonadditive four-body potential, but even when we conservatively assume that the uncertainty is of the order of the nonadditive four-body contribution itself, it is essentially negligible.

    The values of $C$ and $RT \gamma_\mathrm{a}$ computed for all isotopic
    combinations, the natural isotopic mixture, and the average molar mass as well as the values of
    $D$ for the ${}^{20}$Ne isotope and the average molar mass are reported, together with their
    uncertainties, in the \textcolor{blue}{supplementary material}.


\subsection{Analytical fits}
\label{sec:4.4}


We fitted simple analytical functions for $B(T)$, $C(T)$, and $D(T)$ to the calculated values for $B$ and $\beta_\mathrm{a}$ of Hellmann \emph{et~al.}\cite{Hellmann:2021} and $C$, $D$, and $RT\gamma_\mathrm{a}$ of this work for the natural isotopic composition of neon, using the relations between the density and acoustic virial coefficients given above to incorporate $\beta_\mathrm{a}$ and $RT\gamma_\mathrm{a}$ into the $B(T)$ and $C(T)$ fits. The functions are given by
\begin{equation}
\label{eq:Bfit}
X(T) = \sum_{i=1}^{i_\mathrm{max}(X)} a_{X,i}\left(\frac{T}{100\;\mathrm{K}}\right)^{b_{X,i}},
\end{equation}
with
\begin{equation}
\label{eq:Bfit_2}
b_{X,i} = c_{X,0} + c_{X,1}i + c_{X,2}i^2,
\end{equation}
where $X=B,C,D$, $i_\mathrm{max}(B)=10$, $i_\mathrm{max}(C)=i_\mathrm{max}(D)=11$, and $a_{X,i}$, $c_{X,0}$, $c_{X,1}$, and $c_{X,2}$ are the independent fit parameters. We first fitted $B(T)$ simultaneously to the $B$ and $\beta_\mathrm{a}$ values of Hellmann \emph{et~al.}, which we weighted by the squared inverses of their combined uncertainties. In the next step, we fitted $C(T)$ simultaneously to the $C$ and $RT\gamma_\mathrm{a}$ values of this work, using the $B(T)$ fit to calculate $B$ and its first and second temperature derivative in the expression for $RT\gamma_\mathrm{a}$. Because we did not perform PIMC calculations of the fourth acoustic virial coefficient $(RT)^2\delta_\mathrm{a}$, we fitted $D(T)$ only to our calculated $D$ values. The calculated values for $C$, $D$, and $RT\gamma_\mathrm{a}$ were weighted in the fits by the squared inverses of their statistical uncertainties. From the $B(T)$, $C(T)$, and $D(T)$ fits, $(RT)^2\delta_\mathrm{a}$ values can finally be derived, but unfortunately it is not possible to determine their uncertainties in a straightforward manner. Therefore, we do not provide uncertainty estimates for $(RT)^2\delta_\mathrm{a}$. We note that the exponents $b_{X,i}$ resulting from the fits are always negative, so that the extrapolation behavior of the analytical $B(T)$, $C(T)$, and $D(T)$ functions to temperatures above 5000\;K is physically reasonable.

We also provide analytical fits of the combined standard uncertainties $u(X)$ with $X=B,C,D,\beta_\mathrm{a},RT\gamma_\mathrm{a}$ for natural neon. The functional form is similar to that used for $B(T)$, $C(T)$, and $D(T)$:
\begin{equation}
\label{eq:uBfit}
u\left[X(T)\right] = \sum_{i=1}^{i_\mathrm{max}\left[u(X)\right]} a_{u(X),i}\left(\frac{T}{100\;\mathrm{K}}\right)^{b_{u(X),i}},
\end{equation}
with
\begin{equation}
\label{eq:uBfit_2}
b_{u(X),i} = c_{u(X),0} + c_{u(X),1}i + c_{u(X),2}i^2,
\end{equation}
where $i_\mathrm{max}\left[u(B)\right]=i_\mathrm{max}\left[u(C)\right]=i_\mathrm{max}\left[u(RT\gamma_\mathrm{a})\right]=4$ and $i_\mathrm{max}\left[u(\beta_\mathrm{a})\right]=i_\mathrm{max}\left[u(D)\right]=5$.

The values of $a_{X,i}$, $b_{X,i}$, $a_{u(X),i}$, and $b_{u(X),i}$ are provided in the \textcolor{blue}{supplementary material}.


\section{Discussion}
\label{sec:5}



\subsection{Third and fourth density virial coefficients}
\label{sec:5.1}

\begin{figure*}
\includegraphics{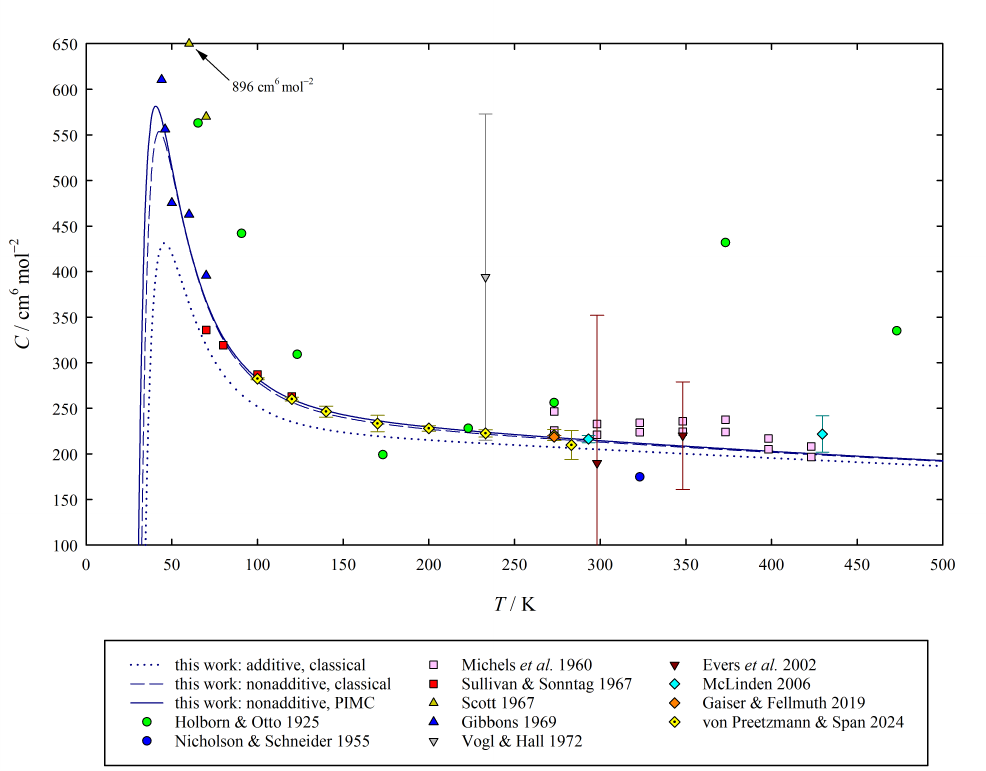}
\caption{Third density virial coefficient $C$ of this work at different levels of theory and experimental data for $C$.\cite{Holborn:1925,Nicholson:1955,Michels:1960,Sullivan:1967,Scott:1967,Gibbons:1969,Vogl:1972,Evers:2002,McLinden:2006,Gaiser:2019,Preetzmann:2024}}
\label{fig:C_abs}
\end{figure*}

Figure~\ref{fig:C_abs}
shows the values for the third density virial coefficient $C$ of natural neon calculated at three different levels of sophistication, namely, classically without the nonadditive three-body potential (i.e., assuming pairwise-additive interactions), classically with the nonadditive three-body potential, and quantum-mechanically by PIMC with the nonadditive three-body potential. It can be seen that the effect of nonadditive three-body interactions is substantial and much larger than the contribution of quantum effects.

Also shown in Fig.~\ref{fig:C_abs} are experimental data for $C$.\cite{Holborn:1925,Nicholson:1955,Michels:1960,Sullivan:1967,Scott:1967,Gibbons:1969,Vogl:1972,Evers:2002,McLinden:2006,Gaiser:2019,Preetzmann:2024} The data of Holborn and Otto\cite{Holborn:1925} were taken from the compilation of Dymond \emph{et~al.}\cite{Dymond:2002} We reanalyzed the value for $C$ (along with the value for $B$) extracted by Gaiser and Fellmuth\cite{Gaiser:2019} at 273.16\;K from dielectric-constant gas thermometry measurements using the most recent first-principles values for the second and third dielectric virial coefficients,\cite{Hellmann:2021,Garberoglio:2021} which were not yet available when Gaiser and Fellmuth analyzed their measurements. The standard uncertainties for the reanalyzed $B$ and $C$ values were kept unchanged from those given by Gaiser and Fellmuth. The papers of Evers \emph{et~al.},\cite{Evers:2002} McLinden,\cite{McLinden:2006} and von Preetzmann and Span\cite{Preetzmann:2024} reported $\left(p,\rho,T\right)$ data measured with single- or dual-sinker densimeters, but the authors of these three papers did not extract virial coefficients from their data. To obtain these coefficients, we performed isothermal fits of the data using a sixth-order virial expansion of the form of Eq.~\eqref{eq:Virial_expansion_density}. In the case of the paper of McLinden, the data first had to be slightly adjusted to lie exactly on isotherms. Only $B$, $C$, and the zero-density limit of $p/(\rho_\mathrm{m}RT)$ were treated as adjustable fit parameters. The values for $D$ were fixed at those obtained from the PIMC calculations, and the fifth and sixth density virial coefficients $E$ and $F$ were fixed at values calculated semiclassically with the pair potential of Hellmann \emph{et~al.}\cite{Hellmann:2021} and the nonadditive three-body potential of this work. A discussion of $E$ and $F$ is beyond the scope of this paper and will be part of a separate publication.\cite{Marienhagen:2026} The standard uncertainties of the resulting $B$ and $C$ values were roughly estimated as the maximum changes in these coefficients (rounded up to the next multiple of $0.01\;\mathrm{cm}^3/\mathrm{mol}$ and $1\;\mathrm{cm}^6/\mathrm{mol}^2$, respectively) when excluding the highest and highest two nominal pressures, except for the $C$ values derived from the measurements of McLinden, for which only the highest nominal pressure was excluded. The uncertainties obtained in this manner are almost always larger and appear to be more realistic than the standard errors obtained from the isothermal fits. In Table~\ref{tab:Virial_exp},
\begin{table}
\footnotesize
\caption{\label{tab:Virial_exp} Values for the second density virial coefficient $B$ and its estimated standard uncertainty $u(B)$ (both in cm$^3$/mol) and for the third density virial coefficient $C$ and its estimated standard uncertainty $u(C)$ (both in cm$^6$/mol$^2$) obtained in this work from the $\left(p,\rho,T\right)$ data of Evers \emph{et~al.},\cite{Evers:2002} McLinden,\cite{McLinden:2006} and von Preetzmann and Span\cite{Preetzmann:2024} and from a reanalysis of the values for $B$ and $C$ derived by Gaiser and Fellmuth\cite{Gaiser:2019} from dielectric-constant gas thermometry measurements.} 
\begin{ruledtabular}
\begin{tabular}{d{3.2}d{2.4}d{1.4}d{3.1}d{3.2}}
\multicolumn{1}{c}{$T/\mathrm{K}$} &
\multicolumn{1}{c}{$B$} &
\multicolumn{1}{c}{$u\left(B\right)$} &
\multicolumn{1}{c}{$C$} &
\multicolumn{1}{c}{$u\left(C\right)$}
\\[0.5ex]
\hline
\\[-1.5ex]
\multicolumn{5}{l}{Evers \emph{et~al.}\cite{Evers:2002}} \\[1ex]
298.15 &   11.73 & 0.46   & 190 & 162   \\
348.15 &   12.29 & 0.15   & 220 &  59   \\[1ex]
\multicolumn{5}{l}{McLinden\cite{McLinden:2006}} \\[1ex]
293.15 &  11.462 & 0.02   & 216.3 &  4  \\
429.75 &  13.118 & 0.10   & 221.8 & 20  \\[1ex]
\multicolumn{5}{l}{Gaiser and Fellmuth\cite{Gaiser:2019}} \\[1ex]
273.16 & 11.0132 & 0.0028\textsuperscript{\textcolor{blue}{a}} & 218.6 & 1.5\textsuperscript{\textcolor{blue}{a}} \\[1ex]
\multicolumn{5}{l}{von Preetzmann and Span\cite{Preetzmann:2024}} \\[1ex]
100.00 &  -4.327 & 0.01   & 282.4 &  1  \\
120.00 &   0.174 & 0.02   & 260.0 &  2  \\
140.00 &   3.230 & 0.03   & 246.3 &  6  \\
170.00 &   6.290 & 0.05   & 233.3 &  9  \\
200.00 &   8.273 & 0.01   & 227.9 &  3  \\
233.15 &   9.771 & 0.02   & 222.5 &  4  \\
273.15 &  10.973 & 0.02   & 220.6 &  6  \\
283.15 &  11.278 & 0.05   & 209.7 & 16 
\end{tabular}
\end{ruledtabular}
\flushleft{\textsuperscript{\textcolor{blue}{a}}The uncertainties $u\left(B\right)$ and $u\left(C\right)$ for the reanalyzed $B$ and $C$ values of Gaiser and Fellmuth were kept unchanged from those given in the original paper.}
\end{table}
we provide all values for $B$ and $C$, along with their standard uncertainties, derived in this work from the experimental literature. The values obtained for $B$ from the measurements of Evers \emph{et~al.}, McLinden, and von Preetzmann and Span agree with the calculated values of Hellmann \emph{et~al.}\cite{Hellmann:2021} within $0.15$, $0.06$, and $0.05\;\mathrm{cm}^3/\mathrm{mol}$, respectively. The reanalyzed value for $B$ of Gaiser and Fellmuth differs by only $-0.006\;\mathrm{cm}^3/\mathrm{mol}$ from the respective calculated value of Hellmann \emph{et~al.}
 
\begin{figure}
\includegraphics{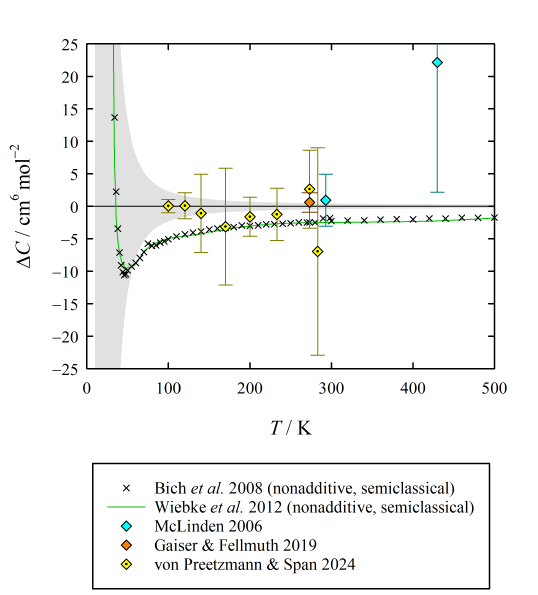}
\caption{Deviations of values for the third density virial coefficient $C$ calculated by Bich \emph{et~al.}\cite{Bich:2008a} and Wiebke \emph{et~al.}\cite{Wiebke:2012} and derived from experiment\cite{McLinden:2006,Gaiser:2019,Preetzmann:2024} from the corresponding PIMC values of this work, whose standard uncertainty is indicated by the gray shading.}
\label{fig:C_diff}
\end{figure}

Most of the experimental data sets for $C$ shown in Fig.~\ref{fig:C_abs} deviate significantly from the PIMC values and from each other and have either no or very large estimated uncertainties. These data are, thus, of little value for validating the PIMC values. However, one of the two data points derived from the measurements of McLinden,\cite{McLinden:2006} the reanalyzed datum of Gaiser and Fellmuth,\cite{Gaiser:2019} and most of the data points derived from the measurements of von Preetzmann and Span\cite{Preetzmann:2024} agree exceptionally well with the calculated values. In Fig.~\ref{fig:C_diff}, 
this agreement is demonstrated more clearly in the form of a deviation plot, in which the standard uncertainties of the PIMC values are indicated by the gray shading. The von Preetzmann and Span values at the lowest two temperatures agree essentially exactly with the PIMC values, with deviations of less than $0.1\;\mathrm{cm}^6/\mathrm{mol}^2$. These two values are also the only ones from all available data sets that have estimated standard uncertainties that are competitive with those of the corresponding PIMC values. Even the very accurate reanalyzed $C$ value of Gaiser and Fellmuth\cite{Gaiser:2019} at 273.16\;K, $(218.6\pm 1.5)\;\mathrm{cm}^6/\mathrm{mol}^2$, does not reach the uncertainty level of the corresponding PIMC value, $(218.0\pm 0.4)\;\mathrm{cm}^6/\mathrm{mol}^2$.

Figure~\ref{fig:C_diff} additionally shows the deviations of the two previous sets of first-principles values for $C$ of Bich \emph{et~al.}\cite{Bich:2008a} and Wiebke \emph{et~al.}\cite{Wiebke:2012} from our PIMC values. In both studies, the calculations were performed semiclassically using an older pair potential of Hellmann \emph{et~al.},\cite{Hellmann:2008a} which Bich \emph{et~al.}\ complemented by the simple ATM potential and Wiebke \emph{et~al.}\ by an extended ATM potential fitted to \emph{ab initio} calculated nonadditive three-body interaction energies for equilateral triangular configurations.\cite{Schwerdtfeger:2009} The $C$ values of Bich \emph{et~al.}\ and Wiebke \emph{et~al.}\ are very similar and overall less consistent with the experimental data shown in the figure than the calculated values of this work, with particularly the data points derived from the measurements of von Preetzmann and Span\cite{Preetzmann:2024} at the two lowest temperatures and the reanalyzed datum of Gaiser and Fellmuth\cite{Gaiser:2019} at 273.16\;K being strong discriminators. The reduction in the deviations from these data achieved by the present calculations stems mostly from the improved treatment of the nonadditive three-body interactions.

\begin{figure}
\includegraphics{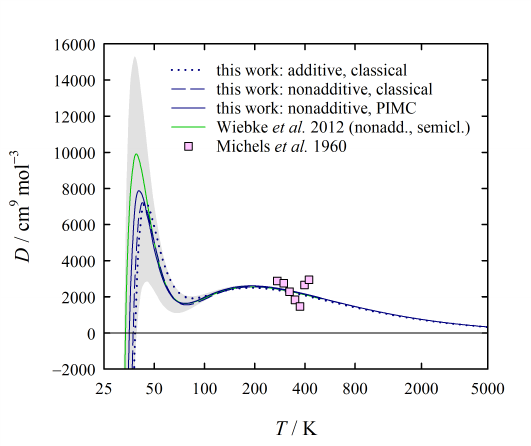}
\caption{Fourth density virial coefficient $D$ of this work at different levels of theory, $D$ values calculated by Wiebke \emph{et~al.},\cite{Wiebke:2012} and the experimental data for $D$ of Michels \emph{et~al.}\cite{Michels:1960} The gray shading indicates the standard uncertainty of the PIMC values.}
\label{fig:D_abs}
\end{figure}

Figure~\ref{fig:D_abs}
shows the values for the fourth density virial coefficient $D$ of natural neon calculated classically without the nonadditive three-body and four-body potentials, classically with the nonadditive three-body and four-body potentials, and quantum-mechanically by PIMC with the nonadditive three-body and four-body potentials. The nonadditive three-body and four-body contributions are not shown separately because the latter is roughly two orders of magnitude smaller than the former.

The only available literature data for $D$ are the calculated values of Wiebke \emph{et~al.}\cite{Wiebke:2012} and the experimental data of Michels \emph{et~al.},\cite{Michels:1960} which are also shown in Fig.~\ref{fig:D_abs}. Wiebke \emph{et~al.}\ obtained their values for $D$, which agree reasonably well with our values, semiclassically with the same pair potential and nonadditive three-body potential as their values for $C$. They also investigated the influence of nonadditive four-body interactions using the simple Bade potential\cite{Bade:1958} and found them, as in the case of our improved calculation, to be very small (their calculated values for this contribution have the wrong sign though). The experimental data of Michels \emph{et~al.}\ exhibit a large scatter, which is not surprising because it is well known that the coefficients of virial expansions fitted to experimental data become more and more correlated with increasing order of the coefficients and thus lose their physical meaning.


\subsection{Third and fourth acoustic virial coefficients}
\label{sec:5.2}


The only experimental data for the third and fourth acoustic virial coefficients $RT\gamma_\mathrm{a}$ and $(RT)^2\delta_\mathrm{a}$ of neon we are aware of are those of Dietl \emph{et~al.}\cite{Dietl:2024} They extracted these coefficients from their highly accurate measurements of the speed of sound (with a standard uncertainty of only 0.0035\%) at temperatures from 200 to 420\;K and pressures up to 100\;MPa. The values for $\beta_\mathrm{a}$ were fixed at those calculated by Hellmann \emph{et~al.}\cite{Hellmann:2021} because the lowest pressures investigated in the experiments (10\;MPa at the two lowest temperatures, otherwise 20\;MPa) were already too high to fit $\beta_\mathrm{a}$ as well.

\begin{figure*}
\includegraphics{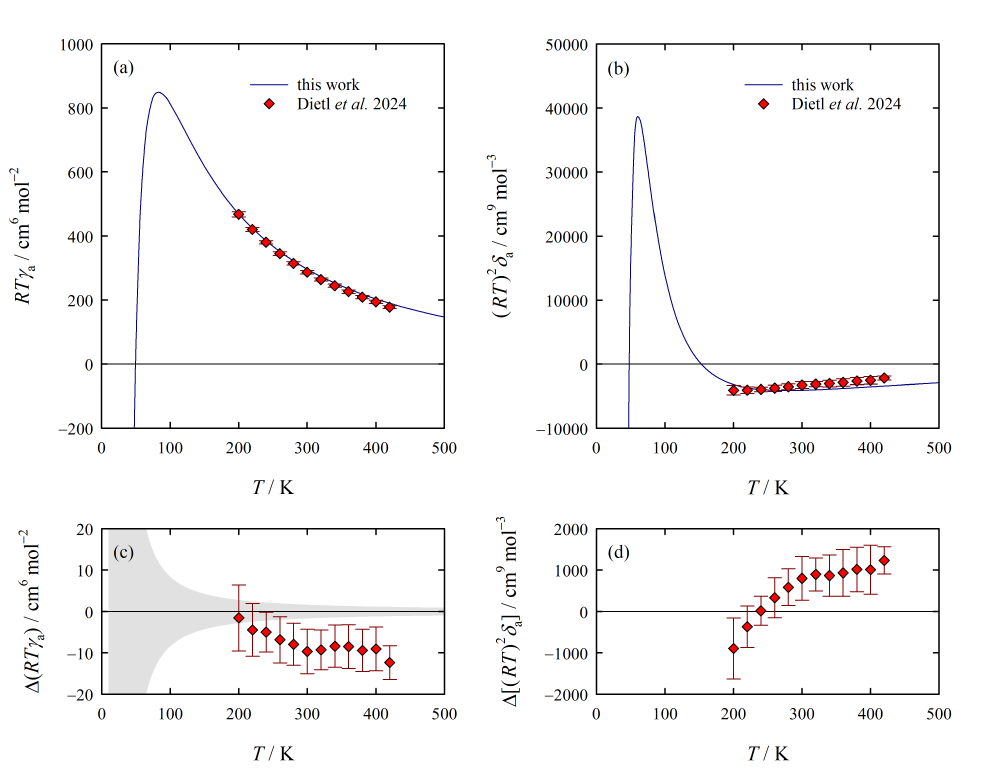}
\caption{(a) Third acoustic virial coefficient $RT\gamma_\mathrm{a}$ of this work at the PIMC level in comparison with the experimental $RT\gamma_\mathrm{a}$ data of Dietl \emph{et~al.}\cite{Dietl:2024} (b) Fourth acoustic virial coefficient $(RT)^2\delta_\mathrm{a}$ derived in this work from the fitted functions for the density virial coefficients in comparison with the experimental $(RT)^2\delta_\mathrm{a}$ data of Dietl \emph{et~al.}\cite{Dietl:2024} (c) Deviations of the $RT\gamma_\mathrm{a}$ data of Dietl \emph{et~al.}\cite{Dietl:2024} from the respective values of this work, whose standard uncertainty is indicated by the gray shading. (d) Deviations of the $(RT)^2\delta_\mathrm{a}$ data of Dietl \emph{et~al.}\cite{Dietl:2024} from the respective values of this work.}
\label{fig:Acoustic}
\end{figure*}

Figures~\ref{fig:Acoustic}(a) and \ref{fig:Acoustic}(b)
show our PIMC values and the data of Dietl \emph{et~al.}\ for $RT\gamma_\mathrm{a}$ and $(RT)^2\delta_\mathrm{a}$, respectively, with Figs.~\ref{fig:Acoustic}(c) and \ref{fig:Acoustic}(d) depicting the deviations of Dietl \emph{et~al.}'s $RT\gamma_\mathrm{a}$ and $(RT)^2\delta_\mathrm{a}$ data from our values. For most temperatures, the agreement for both $RT\gamma_\mathrm{a}$ and $(RT)^2\delta_\mathrm{a}$ is not within the experimental standard uncertainty but still within an expanded uncertainty with $k=2$. A comparison of the patterns of the deviations seen in Figs.~\ref{fig:Acoustic}(c) and \ref{fig:Acoustic}(d) shows that there is a strong negative correlation between the experimental $RT\gamma_\mathrm{a}$ and $(RT)^2\delta_\mathrm{a}$ values, without which their uncertainties would be considerably smaller.


\section{Conclusions}
\label{sec:6}


The third and fourth density virial coefficients $C$ and $D$ and the third and fourth acoustic virial coefficients $RT\gamma_\text{a}$ and $(RT)^2\delta_\text{a}$ of neon have been obtained at temperatures from 10 to 5000\;K with unprecedented accuracy from path-integral Monte Carlo (PIMC) calculations using \emph{ab initio} interatomic potentials. The present calculations extend the recent work of Hellmann \emph{et~al.},\cite{Hellmann:2021} who developed a state-of-the-art \emph{ab initio} pair potential for neon and already calculated the second density and acoustic virial coefficients $B$ and $\beta_\text{a}$.

For this work, we reused Hellmann \emph{et~al.}'s pair potential and developed new nonadditive three-body and four-body potentials. The nonadditive three-body potential is based on counterpoise-corrected supermolecular calculations for 2550 configurations of three neon atoms using basis sets of up to sextuple-zeta quality and levels of theory up to \mbox{CCSDT(Q)}. Relativistic effects were accounted for by calculating DPT2 corrections at the \mbox{CCSD(T)} level of theory. The far less important nonadditive four-body potential is based on counterpoise-corrected supermolecular \mbox{CCSD(T)} calculations performed solely for regular tetrahedra of four neon atoms and represented analytically by a modification of the simple Bade potential.\cite{Bade:1958} The new potential functions developed in this work are provided in the \textcolor{blue}{supplementary material} in the form of Fortran~90 codes.

We calculated $C$, $D$, and $RT\gamma_\text{a}$ directly with the PIMC approach. For $C$ and $RT\gamma_\text{a}$, all isotopic combinations were considered and averaged according to the natural composition, while for $D$ the calculations were performed only for the mass of its most abundant isotope ($^{20}$Ne) and for the average molar mass of the naturally occurring mixture. To obtain uncertainty estimates for $C$, $D$, and $RT\gamma_\text{a}$ from the Monte Carlo calculations, we rigorously propagated the uncertainties provided by Hellmann \emph{et~al.}\ for their pair potential and those estimated for our new nonadditive three-body potential. The contribution of the uncertainty of the nonadditive four-body potential to the uncertainty of $D$ was neglected, which is justified because the nonadditive four-body contribution to $D$ is extremely small. We provide the directly calculated values of $C$, $D$, and $RT\gamma_\text{a}$, the parameters of the analytical fits of $B$, $C$, and $D$ [Eqs.~\eqref{eq:Bfit} and \eqref{eq:Bfit_2}], from which the corresponding values for $\beta_\text{a}$, $RT\gamma_\text{a}$, and finally also $(RT)^2\delta_\text{a}$ can be determined using thermodynamic relations, and the parameters of the analytical fits of the standard uncertainties of $B$, $C$, $D$, $\beta_\text{a}$, and $RT\gamma_\text{a}$ [Eqs.~\eqref{eq:uBfit} and \eqref{eq:uBfit_2}] in the \textcolor{blue}{supplementary material}.

Only few of the available experimental data for $C$, $D$, $RT\gamma_\text{a}$, and $(RT)^2\delta_\text{a}$ are accurate enough for at least a rough validation of our calculations. These data include the value for $C$ determined by Gaiser and Fellmuth\cite{Gaiser:2019} at 273.16\;K from dielectric-constant gas thermometry measurements, which we reanalyzed for this work, as well as some of the values for $C$ extracted by us from the $\left(p,\rho,T\right)$ measurements of McLinden\cite{McLinden:2006} and von Preetzmann and Span.\cite{Preetzmann:2024} The agreement with these values is well within the experimental uncertainties, which are, for the most part, still much larger than those of the calculated values. The previous sets of first-principles values for $C$ of Bich \emph{et~al.}\cite{Bich:2008a} and Wiebke \emph{et~al.}\cite{Wiebke:2012} deviate significantly from these experimental data, which is mostly due to the relatively simple treatment of nonadditive three-body interactions in these two studies. The values for $D$ calculated by Wiebke \emph{et~al.}\ agree reasonably well with our values, whereas the only experimental data set for $D$\cite{Michels:1960} exhibits a very large scatter. In the case of the acoustic virial coefficients $RT\gamma_\text{a}$ and $(RT)^2\delta_\text{a}$, there are only the recent experimental data of Dietl \emph{et~al.}\cite{Dietl:2024} with which we can compare our values. The agreement for both $RT\gamma_\text{a}$ and $(RT)^2\delta_\text{a}$ is mostly  within the expanded experimental uncertainty with coverage factor $k=2$.

The virial coefficients calculated in this work should be highly useful for establishing neon as an alternative working gas in temperature and pressure metrology. A significant further reduction of the uncertainties would be desirable at temperatures below 100\;K, but the required improvements to both the pair potential and the nonadditive three-body potential would be difficult to achieve with current computational resources.


\begin{appendix}
\section{Uncertainty propagation for the third acoustic virial coefficient}
\label{sec:appendix}
  In this Appendix, we derive expressions for estimating the uncertainty propagated from
  the potentials to the third acoustic virial coefficient $RT
  \gamma_\mathrm{a}$. We will start from the classical
  expression reported in Ref.~\onlinecite{Garberoglio:2023} and discuss how to extend it to the
  quantum regime, discussing useful approximations.

  First of all, let us point out two typos in Eqs.~(A6) and (A7) of
  Ref.~\onlinecite{Garberoglio:2023}. The first equation should read
  \begin{eqnarray}
      c_{TT} &=& \beta U_3 (\beta U_3 - 2) \e^{-\beta U_3} \nonumber \\
      & & - \sum_{i<j} \beta U_2(r_{ij}) [\beta U_2(r_{ij}) - 2] \e^{-\beta U_2(r_{ij})},
      \label{eq:cTT}
  \end{eqnarray}
  that is, the terms $U_3$ (the nonadditive three-body potential, denoted in this paper as
  $V_{123}$ or, in the compact notation introduced in Sec.~\ref{sec:4.3}, as $V_3$) under the sum
    should be replaced by $U_2(r_{ij}$), which is the pair potential and is 
  denoted here as $V_{12}$ or $V_2$. The second equation, which is the classical expression of $RT
  \gamma_\mathrm{a}$, should read 
  \begin{widetext}
  \begin{eqnarray}
    RT \gamma_\mathrm{a}(T) &=& \frac{8 \pi^2 N_\mathrm{A}^2 }{3} \int \left\{
    \frac{2}{15}b_{TT}(r_{12})b_{TT}(r_{13}) +
    \frac{14}{15} b_T(r_{12}) b_{TT}(r_{13}) +
     b(r_{12}) b_{TT}(r_{13}) \right. \nonumber \\
    & & + \frac{73}{30} b_T(r_{12}) b_T(r_{13}) +    
    \frac{34}{5} b(r_{12}) b_T(r_{13}) +
    \frac{33}{5} b(r_{12}) b(r_{13}) \nonumber \\
    & & - \left. \left[ \frac{2}{15}c_{TT}(r_{12},r_{13},r_{23}) +
    \frac{16}{15}c_T(r_{12},r_{13},r_{23}) +
    \frac{13}{5}c(r_{12},r_{13},r_{23})
    \right] \right\} \dd \Omega_3,
    \label{eq:RTg_classical}
  \end{eqnarray}
  \end{widetext}
  that is, the correct factor in front of the integral is $\pi^2$ instead of $\pi^3$. We refer to
  the original paper for the definition of the quantities appearing in Eq.~(\ref{eq:RTg_classical}).

  The evaluation of the uncertainty proceeds along the lines outlined in Sec.~\ref{sec:4.3} and
  starts by taking the functional derivatives of all quantities appearing in
  Eq.~(\ref{eq:RTg_classical}) with respect to the pair potential $V_2$,
  \begin{eqnarray}
    \frac{\delta b}{\delta V_2} &=& -\beta \e^{-\beta V_2}, \\
    \frac{\delta b_T}{\delta V_2} &=& -\beta (\beta V_2 - 1) \e^{-\beta V_2}, \\
    \frac{\delta b_{TT}}{\delta V_2} &=& -\beta \left[ (\beta V_2)^2 - 4 \beta V_2 + 2
      \right]\e^{-\beta V_2}, \\
    \frac{\delta c}{\delta V_2} &=& -3 \beta \left( \e^{-\beta V_{123}} - e^{-\beta V_2} \right),\\
    \frac{\delta c_T}{\delta V_2} &=& -3 \beta \left[
      (\beta V_{123} - 1) \e^{-\beta V_{123}} -
      (\beta V_2 - 1) e^{-\beta V_2} \right], \\
    \frac{\delta c_{TT}}{\delta V_2} &=& -3 \beta \left[
      (\beta V_{123})^2  -4 \beta V_{123} + 2 \right] \e^{-\beta V_{123}} \nonumber \\
      & &
      - \left[ (\beta V_2)^2  -4 \beta V_{2} + 2 \right] \e^{-\beta V_{2}},
  \end{eqnarray}
  and the nonadditive three-body potential $V_{3}$, 
  \begin{eqnarray}
    \frac{\delta c}{\delta V_{3}} &=& -\beta \e^{-\beta V_{3}}, \\
    \frac{\delta c_T}{\delta V_{3}} &=& -\beta (\beta V_{3} - 1) \e^{-\beta V_{3}}, \\
    \frac{\delta c_{TT}}{\delta V_{3}} &=& -\beta \left[
      (\beta V_{3})^2 - 4 \beta V_{3} + 2 \right] \e^{-\beta V_{3}}.
  \end{eqnarray}
  Collecting all the terms, one arrives at the following expression for the uncertainty on $RT
  \gamma_\mathrm{a}$ propagated from the pair potential:
  \begin{eqnarray}
    u_2(RT \gamma_\mathrm{a}) &=& \frac{8 \pi^2 N_\mathrm{A}^2}{3} \int  \delta V_2 \left|
    \frac{\delta b}{\delta V_2} \left( \frac{66}{5} b + \frac{34}{5} b_T + b_{TT} \right) \right. \nonumber \\
    & & + \frac{\delta b_T}{\delta V_2} \left( \frac{34}{5} b + \frac{73}{15} b_T + \frac{14}{15} b_{TT} \right) \nonumber \\
    & & + \frac{\delta b_{TT}}{\delta V_2} \left( b + \frac{14}{15}b_T + \frac{4}{15} b_{TT} \right) \nonumber \\
    & & - \left. \left( \frac{13}{5} \frac{\delta c}{\delta V_2} +
    \frac{16}{15} \frac{\delta c_T}{\delta V_2} + \frac{2}{15}\frac{\delta c_{TT}}{\delta V_2}
    \right) \right|  \dd \Omega_3,
  \end{eqnarray}
  with the expression for the uncertainty propagated from the nonadditive three-body potential being
  \begin{eqnarray}
    u_3(RT \gamma_\mathrm{a}) &=& \frac{8 \pi^2 N_\mathrm{A}^2}{3} \int  \delta V_{3} \left|
    \frac{13}{5}  \frac{\delta c}{\delta V_{3}} +
    \frac{16}{15} \frac{\delta c_T}{\delta V_{3}} \right. \nonumber \\
    & & + \left. \frac{2}{15}  \frac{\delta c_{TT}}{\delta V_{3}} \right|  \dd \Omega_3.
  \end{eqnarray}
  
  In order to extend this classical result to the quantum case, one can perfom the substitutions $V_2 \to
  \overline{V_2}$ and $V_{3} \to \overline{V_{3}}$ [see Eq.~(\ref{eq:overline_V})]. This
  approach neglects the contribution to 
  derivatives with respect to temperature in the definition of $RT \gamma_\mathrm{a}$ that emerge from
  differentiating the ring-polymer distribution function. However, the resulting contribution is
  generally small as shown in Ref.~\onlinecite{Binosi:2024}, where the approximation outlined here is
  compared to an independent evaluation of the uncertainty of $RT \gamma_\mathrm{a}$ for
  helium, yielding a very good agreement.
\end{appendix}      


\section*{Supplementary material}
\label{sec:7}


See the \textcolor{blue}{supplementary material} for detailed results of the quantum-chemical \emph{ab initio} calculations, Fortran~90 codes for computing the nonadditive three-body and four-body potentials, a table with the corrected values of the second acoustic virial coefficient $\beta_\text{a}$, tables providing the virial coefficients $C$, $D$, and $RT\gamma_\text{a}$ calculated using PIMC, and tables with the parameters of the analytical fits of $B$, $C$, and $D$ and the standard uncertainties of $B$, $C$, $D$, $\beta_\text{a}$, and $RT\gamma_\text{a}$.


\section*{Acknowledgments}
\label{sec:8}


We acknowledge support from Real-K (Project No.\ 18SIB02), which received funding from the EMPIR program co-financed by the Participating States and from the European Union's Horizon 2020 research and innovation program.
G.G. acknowledges the University of Trento for a generous allocation of computational
  resources on its high-performance computing cluster. We thank Dr.\ Allan Harvey (NIST) for helpful discussions and comments on the manuscript.

\section*{Author declarations}
\label{sec:9}


\subsection*{Conflict of Interest}

The authors have no conflicts to disclose.

\subsection*{Author Contributions}

\noindent\textbf{Robert Hellmann}: Conceptualization (equal); Data curation (equal); Formal analysis (equal); Investigation (equal); Methodology (equal); Software (equal); Validation (lead); Visualization (lead); Writing -- original draft (lead); Writing -- review \& editing (equal). \textbf{Giovanni Garberoglio}: Conceptualization (equal); Data curation (equal); Formal analysis (equal); Investigation (equal); Methodology (equal); Software (equal); Validation (equal); Writing -- original draft (equal); Writing -- review \& editing (equal).



\section*{Data availability statement}
\label{sec:10}


The data that support the findings of this study are available within the article and its \textcolor{blue}{supplementary material}.

%

\end{document}